\magnification1200

\rightline{KCL-MTH-14-05}

\vskip 2cm
\centerline
{\bf    Generalised Space-time and Gauge Transformations}
\vskip 1cm
\centerline{ Peter West}
\centerline{Department of Mathematics}
\centerline{King's College, London WC2R 2LS, UK}
\vskip 2cm
\leftline{\sl Abstract}
We consider the generalised space-time introduced by the author in 2003 in the context of the non-linear realisation of the semi-direct product of $E_{11}$ and its first fundamental representation. For all the fields we propose  gauge transformations  which are compatible with the underlying $E_{11}$ structure. A crucial role is played by the generalised vielbein that the generalised space-time possess. We work out the explicit form of  the gauge transformations, at low levels,  in four, five and eleven dimensions. 
\vskip2cm
\noindent

\vskip .5cm

\vfill
\eject

\medskip
 {\bf {1 Introduction}}
\medskip
It has been proposed that the effective low energy theory of strings and branes possess a very large Kac-Moody symmetry called $E_{11}$ [1]. The $E_{11}$ symmetry was encoded in the form of a non-linear realisation. In this  paper space-time was introduced by extending the $E_{11}$ algebra to include the space-time translation generator in an adhoc manner. However, in 2003  it was proposed to take the non-linear of the semi-direct product of $E_{11}$ with its first fundamental representation $l_1$,  denoted by $E_{11}\otimes_s l_1$ [2]. We recall
that the notion of a semi-direct product is well known to physicists as
the Poincar\'e group is just the semi-direct product of the Lorentz group
and the space-time translations which is similarly denoted by $SO(1,D-1)\otimes_s T^D$ where $T^D$ are the translation group in $D$ dimensions.  The theory of the non-linear realisations  of a group $G$ with local subgroup $H$  not only specifies the field and space-time content of the theory but also provides a method to construct the dynamics. For the non-linear realisation of $E_{11}\otimes_s l_1$   we find  a theory with an infinite number of fields associated with $E_{11}$ and a generalised space-time associated with the first fundamental representation $l_1$ which is automatically equipped with a   generalised vielbein and corresponding
generalised tangent space. Examples of 
non-linear realisations can be found in [3,4,1] and  a review of
non-linear realisations, and the $E_{11}$ programme,  can be found in the
book of reference [5]. 
\par
The fields and coordinates can be classified in terms of a level which takes integer values.  This has a technical definition in terms of the underlying $E_{11}$ Kac-Moody algebra which  depends on the dimension. For the  fields  that arise from the non-linear realisation, in less than eleven dimensions,  this definition is equivalent to   defining the level as the number of down  minus up  space-time indices that they carry  and it is the same for the coordinates except that for convenience  we also add plus one. In eleven dimensions it is the same except that we divide the result by three. 
The low level fields are those of the maximal supergravity theories and there is very good evidence for the presence of many of the fields that are found at higher levels in an underlying theory of strings and branes, that is, beyond the supergravity approximation.   The lowest level elements in the $l_1$ representation are the usual space-time translations and so the lowest level coordinates are  those of the usual space-time. However,   the physical meaning of the higher level coordinates is not well understood. 
\par
The Dynkin diagram of $E_{11}$ is given by 
$$
\matrix{
& & & & &&& &\bullet &11&&&
\cr & & & &&& & &| & && &
\cr
\bullet&-&\bullet&-&\ldots &- &\bullet&-&\bullet&-&\bullet&-&\bullet
\cr
1& &2& & & &7& &8& & 9&
&10\cr}
$$
The non-linear realisation of $E_{11}\otimes_s l_1$ leads to theories in  dimensions  eleven and less. They emerge  by taking  different decompositions of $E_{11}$ with respect to different subalgebras.  In particular if we delete node $D$ in the above $E_{11}$ Dynkin diagram and decompose  the $E_{11}$ algebra,  and the $l_1$ representation< 
in terms of the remaining 
$A_{D-1}\otimes E_{11-D}$ subalgebra [6,7,8,9] we find the  theory in $D$ dimensions.
 It follows that the $E_{11}\otimes_s l_1$ non-linear realisation in a given  dimension is equivalent to that  in any other dimension by simply redefining the fields and coordinates as they occur in  the different decompositions [7]. 
\par
The $E_{11}$ conjecture states that the non-linear realisation of $E_{11}\otimes_s l_1$ contains the maximal supergravity theories. That is, if one suitably restricts  the fields and coordinates one finds the maximal supergravity theories. The maximal supergravity theories, by virtue of the large amount of supersymmetry they possess,  have been thought to encode all string and brane effects at low energy. However, what the $E_{11}$ conjecture implies  is that the maximal supergravity theories do not encode {\bf all} effects, but these are to  be found in the non-linear realisation of $E_{11}\otimes_s l_1$. Put another way the conjecture is that {\bf the low energy effective action of strings and branes possess an $E_{11}$ symmetry and it is  the non-linear realisation of $E_{11}\otimes_s l_1$. }
\par
As we have mentioned the lowest level coordinates of the $l_1$ representation are  the  usual coordinates of space-time,  $x^a$. In eleven dimensions the $l_1$ representation leads to the coordinates [2,13]
$$
x^a (0), x_{ab}(1), x_{a_1\ldots a_5}(2), x_{a_1\ldots a_7,b}(3), x_{a_1\ldots a_8}(3),
$$
$$
\  x_{\hat b_1\hat b_2 \hat b_3, \hat a_1\ldots \hat a_8}\ (4),
\  x_{(\hat c \hat d ), \hat a_1\ldots \hat a_9}\ (4),
\  x_{\hat c\hat d,\hat a_1\ldots \hat a_9}\ (4),\ 
\  x_{\hat c,\hat a_1\ldots \hat a_{10}}\ (4),\ 
\ldots  
\eqno(1.1)$$
where the number in brackets gives the level. Indeed, reference [13] found the coordinates up to and including level seven. The algebraic equations determining the  generalised coordinates  in lower dimensions were  given in reference [10] and the results applied to three dimensions where the coordinates corresponding to the first two entries in the table below for three dimensions were found. In the $l_1 $ representation at the level beyond the usual coordinates of space-time one finds coordinates which are 
 scalars under the Lorentz group but transform as  
the 10,
$\overline {16}$, 
$\overline {27}$, 56 and
$248\oplus 1$ of SL(5), SO(5,5),
$E_6$. $E_7$ and $E_8$ for $d$ equal to seven, six, five,  four and
three dimensions respectively  [10,11]. 
It is straightforward to compute  the generalised coordinates that are forms, that is, carry completely anti-symmetrised space-time indices which the $l_1$ contains. This 
result for the generators of the $l_1$ representation, appropriate to 
$d$ dimensions, are given in the table [9,10,11,12]. 
\par
The coordinates of the generalised space-time are in one to one correspondence with the generators of the $l_1$ representation and so they can  be easily read off from the above table.  One sees  in
the first column the Lorentz scalar coordinates mentioned above. The  generalised 
tangent space structure that is inherited from the  coordinates is readily apparent  from the table.  
\par 
There is good evidence that the $l_1$ representations also contains all the brane charges  and as a result the way the non-linear realisation is constructed automatically encodes  a one to one correspondence between the coordinates of the
generalised space-time and the brane charges [2,13,12,10]. As such one
can think of each coordinate as associated with a given type of brane
probe.  Furthermore for every
generator  in the Borel
subalgebra  of $E_{11}$ there is at least one  element in the $l_1$
representation [13],  and  as a result   for every field at low level in
the non-linear realisation one finds a corresponding coordinate. This correspondence is apparent at low levels from the above table and the fields at low levels given in references [8], [14]; see page 53 of the former reference for the table giving the $E_n$ representations of all forms.  The papers [10,12,13] also discussed the appearance in the $l_1$ representation of "exotic" representation and so the appearance of  "exotic" brane charges and their associated with coordinates.

\bigskip

 {\centerline{\bf {The generators   in the $l_1$ representation  in D
dimensions}}}
\medskip
$$\halign{\centerline{#} \cr
\vbox{\offinterlineskip
\halign{\strut \vrule \quad \hfil # \hfil\quad &\vrule Ê\quad \hfil #
\hfil\quad &\vrule \hfil # \hfil
&\vrule \hfil # \hfil Ê&\vrule \hfil # \hfil &\vrule \hfil # \hfil &
\vrule \hfil # \hfil &\vrule \hfil # \hfil &\vrule \hfil # \hfil &
\vrule \hfil # \hfil &\vrule#
\cr
\noalign{\hrule}
D&G&$Z$&$Z^{a}$&$Z^{a_1a_2}$&$Z^{a_1\ldots a_{3}}$&$Z^{a_1\ldots a_
{4}}$&$Z^{a_1\ldots a_{5}}$&$Z^{a_1\ldots a_6}$&$Z^{a_1\ldots a_7}$&\cr
\noalign{\hrule}
8&$SL(3)\otimes SL(2)$&$\bf (3,2)$&$\bf (\bar 3,1)$&$\bf (1,2)$&$\bf
(3,1)$&$\bf (\bar 3,2)$&$\bf (1,3)$&$\bf (3,2)$&$\bf (6,1)$&\cr
&&&&&&&$\bf (8,1)$&$\bf (6,2)$&$\bf (18,1)$&\cr Ê&&&&&&&$\bf (1,1)$&&$
\bf
(3,1)$&\cr Ê&&&&&&&&&$\bf (6,1)$&\cr
&&&&&&&&&$\bf (3,3)$&\cr
\noalign{\hrule}
7&$SL(5)$&$\bf 10$&$\bf\bar 5$&$\bf 5$&$\bf \overline {10}$&$\bf 24$&$\bf
40$&$\bf 70$&-&\cr Ê&&&&&&$\bf 1$&$\bf 15$&$\bf 50$&-&\cr
&&&&&&&$\bf 10$&$\bf 45$&-&\cr
&&&&&&&&$\bf 5$&-&\cr
\noalign{\hrule}
6&$SO(5,5)$&$\bf \overline {16}$&$\bf 10$&$\bf 16$&$\bf 45$&$\bf \overline
{144}$&$\bf 320$&-&-&\cr &&&&&$\bf 1$&$\bf 16$&$\bf 126$&-&-&\cr
&&&&&&&$\bf 120$&-&-&\cr
\noalign{\hrule}
5&$E_6$&$\bf\overline { 27}$&$\bf 27$&$\bf 78$&$\bf \overline {351}$&$\bf
1728$&-&-&-&\cr Ê&&&&$\bf 1$&$\bf \overline {27}$&$\bf 351$&-&-&-&\cr
&&&&&&$\bf 27$&-&-&-&\cr
\noalign{\hrule}
4&$E_7$&$\bf 56$&$\bf 133$&$\bf 912$&$\bf 8645$&-&-&-&-&\cr
&&&$\bf 1$&$\bf 56$&$\bf 1539$&-&-&-&-&\cr
&&&&&$\bf 133$&-&-&-&-&\cr
&&&&&$\bf 1$&-&-&-&-&\cr
\noalign{\hrule}
3&$E_8$&$\bf 248$&$\bf 3875$&$\bf 147250$&-&-&-&-&-&\cr
&&$\bf1$&$\bf248$&$\bf 30380$&-&-&-&-&-&\cr
&&&$\bf 1$&$\bf 3875$&-&-&-&-&-&\cr
&&&&$\bf 248$&-&-&-&-&-&\cr
&&&&$\bf 1$&-&-&-&-&-&\cr
\noalign{\hrule}
}}\cr}$$
\medskip

\par
In fact there have been  previous discussions  which considered an  extension of our normal space-time.  In 1990 it was proposed  [15,16] that the first quantised  string  should move in an extended  space-time that had  in addition to the usual coordinates $x^a$ also the  coordinates $y_a$. The motivation was to manifestly encode T duality; the above coordinates belonging to the vector representation of O(D,D). Such an assumption was natural considering the way  that the string on a torus develops an additional zero mode that describes  its ability to wind around the torus.  
 This extension of space-time was also  used  in  1993 to give Êa  field theory  which was manifestly T duality invariant [16]. This involved formulating a field theory that described the massless fields of the NS-NS sector of the superstring in the extended space-time just mentioned. 
\par
The  $E_{11}\otimes_s l_1$ non-linear realisation, introduced in 2003,  has been computed in various  dimensions, at low levels, for the fields and generalised coordinates and in particular including coordinates of the generalised space-time which are in addition to those of the usual space-time.  In 2007  all gauged maximal supergravities in five dimensions we constructed,  for the first time, by taking the fields to depend on some of the generalised coordinates [9]. In 2009 the papers    [18,19] computed the $E_{11}\otimes_s l_1$ non-linear realisation in four dimension keeping the 56 Lorentz scalar coordinates in addition to the usual four coordinates of space-time, but only  the fields at level zero, that is, the metric and the scalar fields. 
\par
More recently a more systematic construction of the $E_{11}\otimes_s l_1$ has been undertaken   in eleven dimensions [20] and four dimensions [21]. In the former paper the eleven-dimensional  equations of motion involving the fields at levels up to and including the dual graviton (level 3) and the generalised  coordinates $x^a$, $x_{a_1a_2}$ and $x_{a_1\ldots a_5}$ were given. While in reference [21], the four dimensional equations involving the scalars and vectors and the coordinates at levels zero and one were given. It was found that once one specified the Lorentz,  and  in the case of four dimensions also the SU(8),   character of the equations they were unique. Furthermore these equations  when restricted to the usual space-time and fields did indeed agree with those of maximal supergravity. However, the equations relating the usual graviton to the dual graviton are the subject of a further study to be published [31]. 
\par
Starting in 2009  significant  number of papers were devoted to the construction of what was called   doubled field theory. This theory is     essentially equivalent to the theory put forward in   the old work of references [15,16]; the equivalence was explained  in reference [22] where one can also find references to this more recent work. Doubled field theory, or more appropriately Siegel theory,   described the massless fields of the NS-NS  sector of the superstring, and it can be viewed as the non-linear realisation of $E_{11}\otimes_s l_1$ in ten  dimensions at level zero [23]. The extension of Siegel theory to include the Ramond-Ramond sector was first given in reference  [24];  it  was  just the  $E_{11}
\otimes_s l_1$ non-linear realisation at level one. 
\par
The goal of the $E_{11}$ programme is to find the symmetries of the effective action of strings and branes and,  as is often correctly stressed,  it is more ambitious that proposals to simply reformulate supergravity theories. However, one usually  considers the $E_{11}\otimes_s l_1$ non-linear realisation at low levels and then one finds the usual fields of the supergravity theories, but also the generalised coordinates mentioned above. As such,  the references given above on  the $E_{11}\otimes_s l_1$  non-linear realisation contain specific  proposals for the fields and generalised space-time,  equipped with a generalised vielbein and tangent spaces, as well as the dynamics for the low level fields. More recently there have been a number of papers on what has become known as "exceptional generalised geometry." The starting point for these papers  are the generalised space-times and tangent spaces, given in previous papers,  of  the $E_{11}\otimes_s l_1$ non-linear realisation. 
For example,  references [25] and [26]   take  generalised space-times which consists of the  first two columns in table one in five and four dimensions. While  references [27] take the tangent spaces of the generalised space-times of the $E_{11}\otimes_s l_1$ non-linear realisation, for example,   it  takes the eleven dimensional tangent space to be that corresponding to the coordinates in equations (1.1) and in the notation of the last paper in reference [27], in particular 
equation (2.16), it is 
given as 
$$
TM\oplus \Lambda^2 T^\star\oplus \Lambda^5T^\star M\oplus (T^\star M\otimes \Lambda^7T^\star M)$$
While the 
$E_{11}\otimes_s l_1$ non-linear realisation specifies the dynamics in these other approaches  the dynamics is hoped to emerge from a generalisation of 
Riemannian geometry to the above  extensions of the usual space-time, or tangent spaces,  following the earlier work of [17] and mathematical development of [28]. 
\par
One puzzle that was apparent from the earliest paper [1] on $E_{11}$ was that although one finds the field of gravity, and   gauge fields,  there was no very obvious, or compelling,  way of introducing the well known local transformations that these field usually possess into the theory. Although the non-linear realisation  contains a local symmetry, which is the  tangent space group of the generalised space-time, its role is to  gauge away the fields in the group element that are associated with the negative roots; indeed at lowest level it is just the local Lorentz group. As such this local symmetry  is not a local symmetry in the sense of diffeomorphisms or gauge transformations. 
In the original  paper, following the approach to gravity of reference [3], the  simultaneous non-linear realisation with the conformal group was taken and this did indeed restrict the constants that arise in the construction of the dynamics in just such a way that the theory was invariant under diffeomorphisms and gauge transformations [1]. Unfortunately there has not so far been found a clear way to extend this method to the full $E_{11}$ algebra. A different approach was taken in references [18,19] mentioned above;  the undetermined coefficients that arise when constructing the dynamics of the $E_{11}\otimes_s l_1$ non-linear realisation  in four dimensions at low levels were fixed by  demanding that the theory be diffeomorphism and gauge invariant. 
\par
It was observed in constructing the  maximally supersymmetric gauged supergravities in five dimensions [9] that the gauge transformations could be viewed as diffeomorphisms of the generalised coordinates   see equations (5.3.35-46). This suggests that one may be able to formulate gauge transformations in a simple way in the $E_{11}\otimes_s l_1$ non-linear realisation if one utilised the generalised space-time it contains. Indeed this is the case and 
in this paper we give a formula that specifies the gauge transformations, including diffeomorphisms,  of all the fields in the non-linear realisation. 
This formula, given, for example,  in equation (3.13), is expressed  in terms of the generalised vielbein which is  automatically  encoded in the $E_{11}\otimes_s l_1$ non-linear realisation in a simple manner. As such the formula makes essential use of not just the coordinates, found at levels zero and one, that is,  the usual coordinates of space-time and those that  are Lorentz scalars (column one of table one), but also the higher level coordinates. In particular the gauge transformation of a field can be thought of as arising from the elements, or equivalently coordinates, that are associated with that field in the $l_1$ representation. The results agree with  the  gauge transformations of the fields in the non-linear realisation that one would usually expect. To give a simple example,  the three form in eleven dimensions is associated in the $l_1$ representation with the two form coordinate and this agrees with the expected two form gauge symmetry of such a field.  
\par
In section two we explain how   general coordinate transformations can be thought to arise in the familiar theory of gravity when it is viewed as a non-linear realisation;  this will prove instructive for the case of interest in this paper. In section three we first review  the aspects of the 
$E_{11}\otimes_s l_1$ non-linear realisation that are required,  and in particular the way it leads to a generalised space-time and vielbein.  We then  discuss the compatibility of the gauge transformations with the underlying $E_{11}$ structure and finally give a formula for the gauge transformations. Section four contains the detailed working out of these gauge transformations  in five, four and eleven dimensions for  some fields of interest at low levels. In section five we further develop the theory behind the gauge transformations and propose explicit forms for the linearised and non-linear gauge transformations.


\medskip 
{\bf 2. Diffeomorphisms in gravity viewed as a  non-linear realisation}
\medskip
In this section we will show how to introduce diffeomorphisms in the context of the formulation of gravity  as a non-linear realisation  of $GL(D)\otimes_s l_1^{SL(D)}$ where $l_1^{SL(D)}$ is the D-dimensional translation group which is  constructed from the  first fundamental representation of GL(D) [3,1]. Although diffeomorphisms in gravity are extremely well known this discussion will prove useful as it has a natural generalisation to introducing local  transformations in  the non-linear realisation of $E_{11}\otimes_s l_1$. 
\par
To construct the non-linear realisation of $GL(D)\otimes_s l_1^{SL(D)}$ we start from the group element 
$g= e^{x^aP_a} e^{h_a{}^b K^a{}_b}$ which is subject to the transformations $g\to g_0gh$ where $g_0\in GL(D)\otimes_s l_1^{SL(D)}$ and $h\in SO(1,D-1)$,  which are rigid and local transformations respectively.  The corresponding Cartan forms ${\cal V}$ are given by 
$$
{\cal V}= g^{-1} dg \equiv dx^\mu (e_\mu{}^a P_a+ G_\mu{}_a{}^b K^a{}_b)= 
  dx^\mu( (e^h)_\mu{}^a P_a + (e^{-1}\partial_\mu e)_a{}^b K^a{}_b)
\eqno(2.1)$$
The field $e_\mu{}^a$ can be interpreted as the vielbein as it transforms on its upper index by a local Lorentz rotation and on its lower index by a change of coordinates. Equations of motion,  or an action, which are  invariant under the above transformations of the $GL(D)\otimes_s l_1^{SL(D)}$ non-linear realisation can readily be constructed but they are contain a few constants  whose values are not determined by this method. 
\par
We take as our starting point that the  local symmetries can be thought of as  elements of the $l_1^{SL(D)}$ representation, that is, the translation generators $P_a$ and so consider the parameter  $\xi^a(x)$ which is a function of space-time. At the linearised level the variation of field $h_a{}^b $ should be constructed from  $\xi^a(x)$ and derivatives of the coordinates, that is,  $\partial_a = {\partial\over \partial x^a}$. The coordinates  transform under GL(D) according to the $l_1$ representation which is easily computed from the transformation on the above group element and one finds  that they transform as 
$\delta x^a= -x^c\Lambda_c{}^a$. As such we take the gauge parameters to transform in the same way, namely,   $\delta \xi^a= -\xi^c\Lambda_c{}^a$. Strictly speaking parameters do not transform, but when taking a closure they inherit a transformation from the transformation of the fields and this   has an equivalent effect. 
The derivatives transform in the contragredient representation  which is readily found to be given by  $\delta (\partial_a )=\Lambda_a{}^b\partial_b$. At the linearised level the local  variation of $h_a{}^b$ should maintain the GL(D) character of this field and the only candidate is 
$$
\delta h_a{}^b= \partial_a \xi^b
\eqno(2.2)$$
\par
To construct the non-linear local variation we consider the variation contained in the object 
$(e^{-1})_a{}^\mu\delta e_\mu{}^b$. As explained below in the next section in the context of the general theory $e_\mu{}^b$ can be thought of as a representation of GL(D) and as a result  it follows that $(e^{-1})_a{}^\mu\delta e_\mu{}^b$  belongs to the Lie algebra of GL(D). In fact, it only transforms under local Lorentz transformations and so it must be equal to a quantity which transforms in the same way. 
Such an equation, which    agrees with equation (3.2) at the linearised level,  is given by 
$$
(e^{-1})_a{}^\mu\delta e_\mu{}^b=(e^{-1})_a{}^\mu (\partial_\mu \xi^b + \omega _{\mu ,} {}^b{}_c \xi^c)\equiv (e^{-1})_a{}^\mu D_\mu\xi^b
\eqno(2.3)$$
where $\xi^c\equiv \xi^\tau e_\tau{}^c$ and 
$\omega_{\mu ,}{}_a{}^b  $ which is  the usual  spin connection. The later is given in terms of the above GL(D) Cartan forms by 
$\omega_a,  {}_{bc}= - G_{b, (ca)}+ G_{c, (ba)} + G_{a, [bc]}$. We can regard equation (2.3) as a covariantised version of equation (2.2). 
\par
The variation of equation (2.3) is nothing but a diffeomorphism and a local Lorentz transformation,  indeed one finds that 
$$
\delta e_\mu{}^a= \xi^\tau\partial_\tau e_\mu{}^a+ \partial_\mu \xi^\tau e_\tau{}^a+ \xi^\tau\omega_{\tau ,}{}^a{}_b e_\mu{}^b
\eqno(2.4)$$
In deriving this equation we have used the well known equation 
$$
\partial_\mu e_\nu{}^a-\partial_\nu e_\mu{}^a+ \omega _{\mu ,}{}^a{}_b e_\nu{}
-\omega _{\mu ,} {}^a{}_b e_\nu{}^b=0
\eqno(2.5)$$
which can also be used to solve for $\omega _\mu {}^a{}_b $. 
\par
It will be instructive to evaluate the commutator of two of the transformations given in equation (2.3). Removing the inverse vielbein factor common to both sides of the equation we find that 
$$
[ \delta_{\xi_1}, \delta_{\xi_2}] e_\mu{}^a = D_\mu (D_\nu (e_\tau {}^a\xi_2^\tau) \xi^\nu_1)+ \delta _{\xi_1} \omega_{\mu ,} {}^a{}_b {\xi_2}^b
-(1\leftrightarrow 2)
$$
$$
= D_\mu (e_\tau{}^a \xi_{comp}^\tau )+ \Lambda_{comp}{}^a{}_b e_\mu{}^b
\eqno(2.6)$$
where $\xi_{comp}^\tau = \xi_1{}^\nu\partial_\nu \xi_2^\tau -\xi_2{}^\nu\partial_\nu \xi_1^\tau$,  $\Lambda_{comp}{}^a{}_b= -3R^{a}{}_b{}_{,cd} \xi_1^c\xi_2^d$ and the Riemann tensor is given as usual, in matrix notation, 
 by 
$ [D_\mu,D_\nu ] V^a = R_{\mu\nu ,}{}^{a}{}_b V^b$ for any vector $V^b$.  We recognise the result as the expected composite general coordinate transformation and a local Lorentz rotation built out of the Riemann tensor. In deriving this result use was made of the standard identity $R_{[\mu\nu ,|}{}^\lambda{}_{|\rho ]}=0$. One can verify that the commutator of a local Lorentz transformation and the transformation of equation (2.3) vanishes.

\medskip 
{\bf 3. Local  transformations and non-linear realisations}
\medskip
In this section we show how to introduce local, or gauge,  transformations into a non-linear realisation which consists of a semi-direct product group, however, for the sake of concreteness  we will apply it to  the non-linear realisation of
$E_{11}\otimes_sl_1$  which is the one of interest to us in this paper. This non-linear realisation is constructed from  a group element
$g\in E_{11}\otimes_sl_1$ which can be written as 
$$
g=g_lg_E 
\eqno(3.1)$$
In this equation   $g_E$ is a group element of $E_{11}$ and so can be written in the form 
$g_E=e^{A\cdot R}$ where $R$ are the generators of $E_{11}$ and $A$ are the fields in the non-realisation, while  $g_l$ is the group element formed from the generators of the $l_1$ representation and so has the form $e^{z^A L_A} $ where $z^A$ are the coordinates and $L_A$  are the generators of the $l_1$ representation. The fields depend on the generalised space-time. The explicit form of these group elements  can be found in earlier papers on $E_{11}$,  for example in dimensions eleven [20], five [9] and four [21]. 
\par
The non-linear realisation is, by definition, invariant under the transformations 
$$
g\to g_0 g, \ \ \ g_0\in E_{11}\otimes _s l_1,\ \ {\rm as \  well \  as} \ \ \ g\to gh, \ \ \ h\in
I_c(E_{11})
\eqno(3.2)$$
The group element $g_0\in E_{11}$ is a rigid transformation, that is, it is  a constant,  while $h$ belongs to the Cartan involution subalgebra of $E_{11}$, denoted $I_c(E_{11})$ and it is a local
transformation, that is,  it depends on the generalised space-time. The latter is the Kac-Moody analogue of the maximal compact of subalgebra for finite dimensional semi-simple Lie groups. 
As the generators in $g_l$ form a representation of $E_{11}$ the above transformations for $g_0\in E_{11}$ can be written as 
$$
g_l\to g_0 g_lg_0^{-1}, g_E\to g_0 g_E\quad {\rm and } \quad g_E\to g_E h
\eqno(3.3)$$
As a consequence the coordinates are inert under the local transformations but transform under the rigid  transformations as 
$$
z^A l_A\to z^{A\prime} l_A=g_0 z^Al_A g_0^{-1} = z^\Pi D(g_0^{-1})_\Pi {}^AL_A
\eqno(3.4)$$
For the differential, and  when  written  in matrix from, this transformation is given by  
$dz^T
\to dz^{T\prime}= dz ^T D(g_0^{-1})$.   The derivative
$\partial_\Pi\equiv {\partial\over \partial z^\Pi}$ in the generalised space-time 
transforms as $\partial_\Pi^\prime= D(g_0)_\Pi{}^\Lambda \partial_\Lambda$. 
\par 
The $l_1$ representation   of  $E_{11}$ is, by definition,  given by 
$$
U(k)( L_A)\equiv k^{-1} L_A k= D(k)_A{}^B L_B, \quad k\in E_{11}
\eqno(3.5)$$ 
where  $D(k)_A{}^B $  is the
matrix representative. As a result we recognise  the matrix $D(g_0^{-1})$ that appears in equation (3.4) as just this representation, although  the indices are labelled according to the role which they will play later in the physical theory that emerges from the non-linear realisation.  
\par
The dynamics of the non-linear realisation is usually constructed from the Cartan forms 
${\cal V}=g^{-1}d g$.  These are inert under the above  rigid $g_0\in E_{11}$ 
transformations, but  transform under the local
transformations as 
$$
{\cal V}\to h^{-1} {\cal V} h+ h^{-1} d h
\eqno(3.6)$$
Hence if we use the Cartan forms as building blocks, the problem of finding field equations which are invariant under the transformations of equation (3.2) 
 reduces to
finding those that is invariant under the local $h$ transformations of the  subalgebra  $I_c(E_{11})$. 
\par
Since the  Cartan forms  belong to the Lie algebra of $E_{11}\otimes_s l_1$ they can be written as  
$$
{\cal V}= {\cal V}_E+{\cal V}_l, \quad {\rm where } \quad
{\cal V}_E=g_E^{-1}dg_E\equiv dz^\Pi G_{\Pi, \underline \alpha} R^{\underline \alpha},
 \eqno(3.7)$$
where the indices  $\underline\alpha, \underline \beta, \ldots $  label the generators of $E_{11}$, and 
 $$
{\cal V}_l= g_E^{-1}(g_l^{-1}dg_l) g_E= g_E^{-1} dz\cdot l g_E\equiv 
dz^\Pi E_\Pi{}^A l_A  
\eqno(3.8)$$
The first part ${\cal V}_E$ belongs to the Lie algebra of $E_{11}$ and it is just the Cartan form for $E_{11}$ while ${\cal V}_l$ is a sum of generators in the $l_1$ representation. We have used the indices $\underline \alpha, \underline \beta, \ldots $ as we have run out of index sets to use, and we will  use the labels $\alpha,  \beta, \ldots $  for the adjoint representations of $E_n$, $n=4,5,6,7,8$. 
\par
While  both ${\cal V}_E$ and ${\cal V}_l$ are invariant under rigid transformations and under local transformations of equation (3.6) they change as 
$$ 
{\cal V}_E\to h^{-1}{\cal V}_E h + h^{-1} d h\quad {\rm and }\quad 
{\cal V}_l\to h^{-1}{\cal V}_l h 
\eqno(3.9)$$
Examining equation (3.8) we recognise 
${ E}_\Pi{}^{A} $ as the representation  matrix $D(g_E)_\Pi{}^A$, and so 
${ E}_\Pi{}^A = D(g_E)_\Pi{}^A$.
\par
Although the 
Cartan form is inert under rigid transformations,  the rigid transformations do  act on
the coordinate differentials contained in the Cartan form and this action  induces  a corresponding  transformation   on the lower index of ${ E_\Pi{}^A}$. On the other hand,   under a local $I_c(E_{11})$ transformation the generalised vielbein ${E}_\Pi{}^{A} $ transforms on its upper index as governed by  equation (3.8) which, using equation (3.5),  we also recognise  as the representation matrix for the subgroup $I_c(E_{11})$. We may summarise these two results as 
$$
{ E}_\Pi{}^{A\prime} =
D(g_0)_\Pi{}^\Lambda { E}_\Lambda{}^{B}D(h)_B{}^A \quad {\rm or}\quad \quad  (E^{-1})_A{}^{\Pi\prime}= D(h^{-1})_A{}^B (E^{-1})_B{}^\Lambda 
D(g_0^{-1})_\Lambda{}^\Pi
\eqno(3.10)$$
 \par
Thus   $E_\Pi{}^A$ transforms
under a local $I_c(E_{11})$ transformation on its $A$ index and by the inverse of the
coordinate transformation on its $\Pi$ index.  These transformations are consistent with the interpretation of ${ E}_\Pi{}^{A}$ as a generalised vielbein of the generalised space-time and associated with the 
generalised tangent space with the tangent group $I_c(E_{11})$.  
\par
 We have already discussed  the $I_c(E_{11})$ local symmetries which are a well known part of a non-linear realisation and are responsible for gauging away fields in the group element that are associated with the negative root generators. We will now introduce new local symmetries that will turn out to contain the  diffeomorphisms of the usual space-time as well as    the  gauge transformations of the  fields. The later  include the usual gauge transformations of the form fields as well as the more complicated gauge transformations of fields with more complicated index structures. 
We take the point of view that these local transformations 
are associated with the $l_1$ representation and  as such we can denote their parameters  by $\Lambda^\Pi$ and we take these to  depend on the generalised space-time. The quantities  $\Lambda^\Pi$  can be thought of as belonging to the $l_1$ representation and so transform under rigid $E_{11}$ transformations in the same way as  the generalised coordinates,  as given in equation (3.4).  
\par
The  linearised gauge transformations of the fields of the non-linear realisation will  involve  the above gauge parameters $\Lambda^\Pi$ and the generalised space-time derivatives,  $\partial_\Pi$ whose rigid $E_{11}$ transformations  were given above. The $l_1$ representation is in fact a lowest weight state with lowest weight state $P_{1}$ where $P_a, \ a=1,\ldots ,D$ are the space-time translation generators in $D$  dimensions. The $\bar l_1$ representation is, by definition,  a highest weight representation with highest weight state $P_{D}$. 
The generalised space-time derivatives transform in the    $\bar l_1$ representations.  
\par
At the linearised level the local, or gauge,  transformations must be given in terms of the  gauge parameters $\Lambda^\Pi$ and the coordinate derivatives of the generalised space-time $\partial_\Pi$ in such a way that  they give an object that transforms in the adjoint representations,  that is, in the same way as do the fields  under rigid  $E_{11}$ transformations. This is possible as  the tensor product of the $l_1$ and the $\bar l_1$ representations does indeed contain the adjoint representation of $E_{11}$. As a result we can write  the {\bf linearised } gauge variations as 
$$
\delta A_{\underline \alpha}= N_{\underline \alpha }{}_\Sigma{}^\Xi \partial_\Xi\Lambda ^\Sigma, \quad {\rm or } \quad \delta A_\Pi{}^\Lambda= N_\Pi{}^\Lambda {}_ \Sigma{}^\Xi \partial_\Xi\Lambda ^\Sigma
\eqno(3.11)$$
  In the second version of this equation we have used the fact that the generators of $E_{11}$ can be expressed in terms of the matrices of the $l_1$ representation and so we can label the fields contained in the non-linear realisation  as $A_\Pi{}^\Lambda$. In this equation $N_{\underline \alpha} {}_ \Sigma{}^\Xi$,  or equivalently $N_\Pi{}^\Lambda{}_ \Sigma{}^\Xi$,  are constants that convert the $l_1\otimes \bar l_1$ representation into the adjoint representation of $E_{11}$.  We note that at the linearised level there is no distinction between world and tangent indices. While it is true that the fields of the non-linear realisation transform under the adjoint representation of $E_{11}$ at the linearised level, it is usual to use the local subalgebra to gauge away the fields associated with the negative roots and once this has been done one requires a compensating transformation in order to preserve the form of the group element when acted on by rigid $E_{11}$ transformations. This compensation complicates the effect of these transformations and should be taken into account when computing the constants $N_\Pi{}^\Lambda{}_ \Sigma{}^\Xi$. 
\par
We  now construct  the full  non-linear local transformations  and we start by considering  the object 
$$
(E^{-1})_A{}^\Pi \delta E_\Pi{}^B
\eqno(3.12)$$
The generalised vielbein $E_\Pi{}^A$ contains all the fields of the non-linear realisation and so specifying the variation of  this quantity is equivalent to specifying what are the variations of the fields contained in the $E_{11}\otimes_s l_1$ non-linear realisation.  As explained above the generalised vielbein can be thought of as  the matrix representation of $E_{11}$ in   the $l_1$ realisation and as a result the above quantity belongs to the Lie algebra of $E_{11}$.   The expression to which the quantity in equation (3.12) is equal  is to be constructed from the generalised covariant derivatives and the gauge parameters, which also belong to the $l_1$ representation. Taking this into account,  we propose that the {\bf non-linear } gauge transformations are given by 
$$
(E^{-1})_A{}^\Pi \delta E_\Pi{}^B= N_A{}^B {}_C{}^D E_D{}^\Pi 
D_\Pi \Lambda ^C
\eqno(3.13)$$
where $ D_\Pi = \partial_\Pi+ \Omega _\Pi $ and $\Omega _\Pi $ is a connection which transforms as ${\cal V}$ does in equation (3.6). We will discuss the connection further  in this section 5. 
The $\Lambda ^A$ in equation  (3.13) transforms under  local $I_c(E_{11})$ transformations as does the upper index on the generalised vielbein $E_\Pi{}^A$,  while $(E^{-1})_D{}^\Pi D_\Pi$ transforms as the lower index of $(E^{-1})_D{}^\Pi$. One can define a parameter with a world index in the usual way by $\Lambda ^A\equiv \Lambda^\Pi E_\Pi{}^A$. The constants $N_A{}^B {}_C{}^D$ ensure  that the left and right-hand sides of the equation transform in the same way;  the objects that appear directly are inert under the $E_{11}$ rigid transformations,  we must demand that the constants transform under $I_c(E_{11})$ transformations as follows 
$$
D(h^{-1})_A{}^E D(h)_F{}^B N_E{}^F {}_C{}^D =N_A{}^B {}_E{}^F D(h)_F{}^D D(h^{-1})_C{}^E
\eqno(3.14)$$
The construction of these constants is just a question of group theory. We can think of equation (3.13) as a covariantised version of equation (3.11); indeed given an allowed contribution at the linearised level object one can use the generalised vielbein to convert the indices to tangent indices and obtain a contribution to the non-linear result. However, we note that the condition of equation (3.14) involves the subgroup $I_c(E_{11})$ of $E_{11}$ and this may allow more solutions.  This consideration is the same as that which occurs when   constructing the equations of motion using the Cartan forms which transform under $I_c(E_{11})$ and in this case  one does find that the additional possibilities are essential.  The constants $N_A{}^B {}_C{}^D$ are straight forward to computed and this is carried out at low levels in four, five and six dimensions in section four. 
\par
Many of the terms in equation (3.13) are not of a familiar type as they contain derivatives with respect to the higher level coordinates of the generalised space-time. However,   by taking the index $A=a$, where $a$ is the index for the usual tangent space,  we will find the more familiar 
gauge transformations.  The way the expected gauge  transformations arise will become  apparent from the worked examples in four, five and eleven dimensions given in section four. 
\par
It is important to realise that the above formula for the local transformations   makes use of the coordinates of  generalised space-time beyond   the usual coordinates of space-time and the additional Lorentz invariant coordinates, found in the first column of table one. Indeed, 
to get even the gauge symmetries that we are familiar with in supergravity theories one must use the coordinates of the generalised space-time at higher levels.  For example,  in eleven dimensions  the three form and six form are associated with the two form and five form coordinates of the generalised space-time and their two form and five form gauge transformations arise from these coordinates.  Furthermore,  the dual graviton $\hat h_{a_1\ldots a_8,b}$ is expected to have two gauge transformations $\Lambda_{a_1\ldots a_8}$ and $\Lambda_{a_1\ldots a_7,b}$
which are associated with the coordinates with the same index structure that occur in the $l_1$ representation at level three. At the next level,  we find the field $A_{a_1\ldots a_9,b_1b_2b_3}$ which should have the gauge transformations $\Lambda_{a_1\ldots a_8, b_1b_2b_3}$ and $\Lambda_{a_1\ldots a_9,b_1b_2}$ and there are indeed associated coordinates in the $l_1$ representation with these indices, see equation (4.3.6). In fact this is true quite generally as it was shown in reference [13] that if we decompose $E_{11}$ and the $l_1$ representation into representations of  SL(11) and  one finds a certain representation of SL(11) in $E_{11}$,  then one finds  in the $l_1$ representations, at the same level,  all the SL(11) representations  which are found from the one in $E_{11}$ by deleting one index in all possible ways. 
\par


\medskip 
{\bf 4. Gauge transformations in four, five and eleven dimensions.  }
\medskip 
In this section we find the explicit form of equation (3.13) in four, five and eleven dimensions at low levels. Using this result we illustrate the formula by finding  the local transformation of the vector in the four five dimensions and the three form in eleven dimensions. 
\medskip 
{\bf 4.1 Gauge transformations in five dimensions}
\medskip

We now illustrate the  discussion of section three for  the  $E_{11}\otimes_sl_1$ non-linear realisation  in  five dimensions. Deleting node five in the $E_{11}$ Dynkin diagram we find the subalgebra  $GL(5)\otimes E_6$ and decomposing $E_{11}$ with respect to  this algebra we find the  positive level the generators 
$$
K^a{}_b (1,0), R^\alpha (78, 0), R^{aM} (\overline {27} ,1) , R^{a_1a_2} _M (27,2), R^{a_1a_2a_3\alpha} (78, 3)\ldots 
\eqno(4.1.1)$$
The first number in the bracket is the dimension of the $E_6$ representations to which they belong   and the second number is their level. The algebra the generators satisfy can be found in references [9,29] and the additional commutators  in appendix A. The corresponding fields in the non-linear realisation are 
$$
h_a{}^b , A_\alpha , A_{aM} , A_{a_1a_2} ^M , A_{a_1a_2a_3\alpha} , \ldots 
\eqno(4.1.2)$$

\par
Decomposing the $l_1$ representation with respect to $GL(5)\otimes E_6$ subalgebra we find that the  lowest level members   are given by [9]
$$
P_a (1,0), Z^N (\overline {27},1), Z^a_N (27,2), Z^{a_1a_2\alpha} (78,3), Z^{a_1a_2} (1,3) ,\ldots 
\eqno(4.1.3)$$
and the corresponding coordinates by 
$$
x^a , x_N , x_a^N , x_{a_1a_2\alpha} , x_{a_1a_2} ,\ldots 
\eqno(4.1.4)$$
We denote the generalised space-time derivatives with respect to these coordinates as 
$$
\partial_a (0), \partial ^N (-1), \partial ^a_N (-2), \partial^{a_1a_2\alpha} (-3), \partial^{a_1a_2} (-3) , \ldots 
\eqno(4.1.5)$$
where the bracket contains the levels of the derivatives. The definitions of the derivatives is as expected, for example, $\partial_a = {\partial\over \partial x^a},\  \partial^{N} = {\partial\over \partial x_{N}}$. 
\par
The Cartan forms that arise in the non-linear realisation $E_{11}\otimes_s l_1$ in five dimensions can be found in [9]. This paper  derived all the gauged supergravities in five dimensions by taking the fields to have a non-trivial dependence on the generalised space-time. 
\par
Following the general procedure given in section three  we introduce the gauge parameters corresponding to the members of the $l_1$ representation: 
$$
\xi^a (0), \Lambda_N (1),  \Lambda_a^N (2),  \Lambda_{a_1a_2\alpha} (3),  \Lambda_{a_1a_2} (3), \ldots 
\eqno(4.1.6)$$
We now write down the linearised gauge transformations of equation (3.11) in five dimensions at low levels.  To do this we write down all possible terms on the right-hand side that   preserve the level and $SL(5)\otimes E_6$ character of the equation:  
$$
\delta h_a{}^b= \partial_a \xi^b+ {1\over 3}  \delta _a^b\partial ^M\Lambda _M - \partial_N^a \Lambda_b{}^N+{2\over 3} \delta_a^b \partial_N^c \Lambda_c{}^N+\ldots , 
\eqno(4.1.7)$$
$$
\delta A_{aM}= \partial_a \Lambda_M - 10 d_{MPQ} \partial^Q\Lambda^Q_a + e_1 (D^\beta)_M{}^N\partial_N^b\Lambda_{ba\beta}+\dots 
\eqno(4.1.8)$$
$$
\delta \phi^\alpha = -6 (D^\alpha)_M{}^R \partial^M\Lambda_R
+6 (D^\alpha)_M{}^R \partial_R^b\Lambda_b{}^M +\dots 
$$
$$
\delta A_{a_1a_2}{}^{M}=\partial_{[a_1} \Lambda_{a_2]}^M +e_2 (D^\beta)_N{}^M\partial^N\Lambda_{ba_1a_2\beta}+\dots 
\eqno(4.1.9)$$
To be pedagogical we should have placed an arbitrary coefficient in front of each term and  determined the value of this coefficient later, however,  in the interest of brevity we have put in the values of the coefficients that we will find below leaving the  two  constants $e_1$ and $e_2$ so far undetermined. The symbol $d_{MPQ}$ is the totally symmetric $E_6$ invariant tensor and  $(D^\beta)_N{}^M$ the representations matrix of the 27 representation, see appendix A. Further properties of these tensors can be found in references [9] and [29]. 
\par
As explained above,  the linearised variations should belong to the adjoint representation of $E_{11}$, generalising the  $SL(5)\otimes E_6$ character of the equation to an $E_{11}$ character.  It is most useful to compute the rigid $E_{11}$ transformations which follow from taking the rigid group element $g_0= e^{a^{aN}R_{aN}}$. We note this is a negative root generator and so this does not preserve the form of the group element usually used in the non-linear realisation.  The gauge parameters transform like the generalised coordinates  which, evaluating equation (3.4), we find to transform as  
$$
\delta x^a= a^{aN} x_N,\quad  \delta x_N= -10 d_{RSN} a^{bR} x_b^S\quad 
\delta x^N_a= -{12} x_{ba\alpha} a^{bR} (D^\alpha)_R{}^N, \ldots 
\eqno(4.1.10)$$
As a result the generalised space-time derivatives transform as 
$$
\delta (\partial_a)= 0,\quad \delta (\partial^N)= -a^{bN}\partial_b,\quad 
\delta (\partial_N^A)= 10a^{aR} d_{RNP} \partial^P, \ldots 
\eqno(4.1.11)$$
\par
The linearised transformations of the fields arising from taking the rigid transformation $g_0= e^{a_{aN}R^{aN}}$ are as follows 
$$
\delta h_b{}^a= a^{aN} A_{bN} +{1\over 3} \delta_b^a a^{cN} A_{cN}, \quad
\delta \phi_\alpha= 6 (D_\alpha)_P{}^Q a^{aP} A_{aQ},\quad 
$$
$$
\delta A_{aN}= 20 d_{RSN} a^{bR} A_{ba}{}^S+(a^{cN}h_c{}^b + 2a^{bM}
+ a^{bP} \phi_\alpha (D^\alpha)_P{}^M)\delta_{M,N}\delta_{a,b}, \quad 
$$
$$
\delta A_{a_1a_2}{}^N= 18 a^{bR}A_{ba_1a_2\alpha}  (D^\alpha)_R{}^N +\ldots 
\eqno(4.1.12)$$
The last term in brackets in $\delta A_{aN}$ arises from the compensation of the group element and this explains the unusual index positions. 
\par
Transforming the right-hand side of equation (4.1.7-9),  using equations (4.1.10) and (4.1.11),  and comparing this with the transformation of the left-hand side, that is, the transformations of the fields of equation (4.1.12),  we find the constants as they are given in equations (4.1.7-9). It is straightforward to compute the remaining constants by including higher level terms.  
\par
We now  compute the gauge transformation of the fields using equation (3.13) for which  we require the generalised vielbein. This  is easily found  using equation (3.8) and  it  is given by 
$$
{ E}= (det e)^{-{1\over 2}}
\left(\matrix {e_\mu{}^a&- e_\mu{}^c A_{\mu N}& - e_\mu{}^c (2A_{cb}{}^N+
d^{RSN} A_{\mu S}A_{cR})\cr
0&N^{\dot M}{}_N& N^{\dot M}{}_Pd^{RPN} A_{bR} \cr
0&0& (e^{-1})_{b}{}^{\mu} (N^{-1})^N{}_{\dot M}\cr}\right)
\eqno(4.1.13)$$
where $N^{\dot M}{}_N$ is the vielbein on the scalar coset space which is computed in the same way as equation (3.8) but using just the scalar fields. 
The  dot placed over an  index means that it is a world rather than a tangent index. 
\par
Taking $A=a$ and $B=M$ we find that the left-hand side of equation (3.13) is given by 
$$
E_a{}^\Pi \delta E_\Pi{}_M= -(e^{-1})_a{}^\mu \delta A_{\mu M} + A_{a\dot N} \delta N^{\dot N}{}_M
= -(e^{-1})_a{}^\mu \delta A_{\mu \dot N} N^{\dot N}{}_M
\eqno(4.1.14)$$
We observe  that it contains the variation of the vector field with world  indices. 
Taking into account equation (4.1.8) we find that the right-hand side of equation (3.13), for the indices we are considering, takes the form 
$$
(E^{-1})_a{}^\Pi D_\Pi \Lambda_M
- 10d_{MPQ} (E^{-1})^P{}^\Pi D_\Pi \Lambda_a{}^Q+\ldots 
\eqno(4.1.15)$$
Using the expression for the generalised vielbein of equation (4.1.13) 
we find that the gauge variation of the vector field is given by 
$$
\delta A_{\mu \dot M} = \tilde D_\mu \Lambda _{\dot M} + A_{\mu \dot P} \tilde D^{\dot P} \Lambda _{\dot M}- 10d_{MPQ}(N^{-1})^P{}_{\dot T} (\tilde D^{\dot T} \Lambda_{\mu}{}^{Q}
-d^{RSQ} \Omega^{\dot T} {}_{,\mu R} \Lambda_S)+\ldots 
\eqno(4.1.16)$$
where $\tilde D_\Pi$ is the covariant derivative with a connection that includes on the level zero contributions and $ \Omega^{\dot T} {}_{,\mu R}=\partial^{\dot T} A_{\mu R}$.  
\par
The equations of motion of the $E_{11}\otimes_s l_1$ non-linear realisation can be computed following  the same techniques that were used to  find them  in eleven [20] and four dimensions [21]. This technique used the Cartan forms and their local transformations, but we now compute the first term in the equation of motion of the vector equation using the rigid transformations computed in this paper. Under the transformations of equation (4.1.12) we find that only the combination 
$$
\partial _{[a_1 }A_{a_2] N}+ 10 d_{NRS} \partial^R A_{a_1a_2}{}^{S}
\eqno(4.1.17)$$
leads to an object with is totally anti-symmetric in three indices which is given by 
$$
-30d_{RSN}a^{bR}\partial_{[b}A_{a_1a_2 ]}{}^S
\eqno(4.1.17)$$
We require an object antisymmetric in three indices as only in this case can we find a term that possess  the same Lorentz character as the other variations which can be collected to give the equation of motion of the scalars.  
\par
We note that the first terms in the vector equation of motion, given in equation (4.1.17) are indeed invariant under the gauge transformation of equation (4.1.7-9) up to the required level. This leads one to hope that the full equations are invariant under the full non-linear gauge transformations.

\medskip 
{\bf 4.2 Gauge transformations in four  dimensions}
\medskip

We now consider the four dimensional theory. Deleting node four in the above $E_{11}$ Dynkin diagram we find the subalgebra $GL(4)\otimes E_{7}$ and decomposing $E_{11}$, and the $l_1$ representation,  into this subalgebra we find the four dimensional theory. In this decomposition the positive level generators of $E_{11}$ are given by [21] 
$$
K^a{}_b (15, 1,0) ,\  R^\alpha (1, 133, 0);\  R^{aN} (4, 56, 1) ;\ 
R^{a_1a_2\alpha} (6, 133, 2),$$
$$
\  \hat K ^{ab} (10, 1, 2) ;\ 
R^{a_1a_2a_3\lambda} (4, 912, 3), \  R^{a_1a_2, b N} (20, 56, 3); \ldots 
\eqno(4.2.1)$$ 
where the  first two figures in the brackets
indicate the dimensions of the SL(4) and
$E_7$ representations respectively, while the last  figure is the level.
In fact it is simpler to also  decompose $E_7$ into its SL(8) subalgebra and work in terms of representations of $GL(4)\otimes SL(8)$. One can then reconstruct  the representations of $E_{7}$ when required. In terms of this decomposition the generators of $E_{11}$ are given by [21]
$$
K^a{}_b (15,1,0), \ R^I{}_J (1,63, 0),\  R^{I_1\ldots I_4} (1, 70, 0), \ 
R^{a}{}^{I_1I_2} (4,28,1),\ R^{a}{}_{I_1I_2} (4, \overline {28},1), 
$$
$$
\ R^{a_1a_2 }{}^I{}_J (6,63,2),\ R^{a_1a_2 }{}^{I_1\ldots I_4} (6,70,2), \ 
\hat K^{a_1a_2} (10, 1, 2),
$$
$$
R^{a_1 a_2, b} {}_{I_1I_2} (20, \overline { 28}, 3),\ R^{a_1 a_2,b } {}^{I_1I_2} (20,  28, 3),\ R^{a_1 a_2a_3} {}_{I_1I_2I_3} {}^{J} (4,  420, 3),\ 
$$
$$
 R^{a_1 a_2 a_3} {}^{I_1I_2I_3} {}_{J}\  (4,  420, 3),\ 
R^{a_1 a_2 a_3} {}_{(I_1I_2)}  (4, 36, 3),\ 
R^{a_1 a_2a_3} {}^{(I_1I_2)}  (4, \overline { 36}, 3),\ldots 
\eqno(4.2.2)$$
The second entry in the brackets denotes the SL(8) representation of the fields. 
The corresponding fields in the $E_{11}\otimes_s l_1$ non-linear realisation are in one to one correspondence with these generators and so are given by 
$$
h_a{}^b , \ \phi^I{}_J,\  \phi_{I_1\ldots I_4} , \ 
A_{a}{}_{I_1I_2} ,\ A_{a}{}^{I_1I_2}, 
\ A_{a_1a_2 }{}^I{}_J ,\ A_{a_1a_2 }{}_{I_1\ldots I_4}, \ 
\hat h_{a_1a_2} ,
$$
$$
A_{a_1 a_2, b} {}^{I_1I_2} ,\ A_{a_1 a_2,b } {}_{I_1I_2} ,\ A_{a_1 a_2a_3} {}^{I_1I_2I_3} {}_{J} ,\ 
$$
$$
 A_{a_1 a_2 a_3} {}_{I_1I_2I_3} {}^{J}\  ,\ 
A_{a_1 a_2 a_3} {}^{(I_1I_2)}  ,\ 
A_{a_1 a_2a_3} {}_{(I_1I_2)}  ,\ldots 
\eqno(4.2.3)$$
We must also  decompose the $l_1$ representation into representations of $GL(4)\otimes SL(8)$ to find [21]
$$
P_a (4,1,0),\  Z^{I_1I_2}  (1, 28, 1), ,\  Z_{I_1I_2}  (1, \overline { 28}, 1),
\ Z^{a}{}^I{}_J (4, 63, 2), \ Z^{a}{}^{I_1\ldots I_4} (4, 70, 2),\ 
Z^a (4,1,2),
$$
$$
Z^{a_1, a_2} {}_{I_1I_2} (16, \overline { 28}, 3),\ Z^{a_1, a_2} {}^{I_1I_2} (16,  28, 3),\ Z^{a_1 a_2} {}_{I_1I_2I_3} {}^{J} (6, \overline { 420}, 3),\ 
 Z^{a_1 a_2} {}^{I_1I_2I_3} {}_{J}\  (6,  420, 3),\ 
$$
$$
Z^{a_1 a_2} {}_{(I_1I_2)}  (6, 36, 3),\ 
Z^{a_1 a_2} {}^{(I_1I_2)}  (6, \overline { 36}, 3),\ldots 
\dots  
\eqno(4.2.4)$$
The generators at level three belong to the $56= 28 + \overline { 28}$ and 
$912= 420+36 +\overline { 420 }+\overline { 36}$-dimensional representations of $E_7$; the decomposition is into representations of SL(8).  The generator $Z^{a_1, a_2} {}_{I_1I_2} $ has no particular symmetry on its $a_1$ and $a_2$ indices. 
\par
In the non-linear realisation these lead to the generalised space-time  of the four dimensional theory which  has 
the  coordinates [21]  
$$
x^a,\  x_{I_1I_2},\  x^{I_1I_2},\ x_{a}{}^I{}_J,\ x_{a}{}^{I_1\ldots I_4},\  \hat x_{a},
$$
$$
x_{a_1, a_2} {}^{I_1I_2} ,\ x_{a_1, a_2} {}_{I_1I_2},\ x_{a_1, a_2} {}^{I_1I_2I_3} {}_{J} ,\ 
x_{a_1, a_2} {}_{I_1I_2I_3} {}^{J}  ,\ 
 $$
$$
x_{a_1, a_2} {}^{(I_1I_2)},\ 
x_{a_1, a_2} {}_{(I_1I_2)}  , \ldots 
\eqno(4.2.5)$$
We denote the generalised derivatives by 
$$
\partial_a (0),\  \partial^{I_1I_2} (-1),\  \partial_{I_1I_2} (-1),\ \partial^{a}{}_I{}^J (-2),\ \partial^{a}{}_{I_1\ldots I_4} (-2),\  \hat \partial^{a} (-2), \ldots 
\eqno(4.2.6)$$
where the number in brackets is the level. They are given in terms of the coordinates in the way expected, for example, 
$\partial_a = {\partial\over \partial x^a},\  \partial^{I_1I_2} = {\partial\over \partial x_{I_1I_2}}$. 
\par
The non-linear realisation of $E_{11}\otimes_s l_1$ was constructed at low levels in reference [21] for the above fields up to the dual graviton and the coordinates at level zero and one. The equations for the scalars and vectors when truncated to the usual fields and coordinates are indeed those of the four dimensional maximal supergravity theory. Furthermore,  once one adopts the expected  $GL(4)\otimes E_{7}$ character of the equation then the rigid symmetries of the non-linear realisation appear to determine these equations uniquely.  
\par
According to the above discussion of section three we should introduce 
local, or gauge, symmetries with  parameters that are in one to one correspondence with the $l_1$ representation and as a result are in one to one correspondence with the coordinates of equation (3.2.4). As a result  the parameters  are  
$$
\xi^a,\  \Lambda_{I_1I_2},\  \Lambda^{I_1I_2},\ \Lambda_{a}{}^I{}_J,\ \Lambda_{a}{}^{I_1\ldots I_4},\  \hat \xi_{a},
$$
$$
\Lambda_{a_1, a_2} {}^{I_1I_2} ,\ \Lambda_{a_1, a_2} {}_{I_1I_2},\ \Lambda_{a_1, a_2} {}^{I_1I_2I_3} {}_{J} ,\ 
\Lambda_{a_1, a_2} {}_{I_1I_2I_3} {}^{J}  ,\ 
 $$
$$
\Lambda_{a_1, a_2} {}^{(I_1I_2)},\ 
\Lambda_{a_1, a_2} {}_{(I_1I_2)}  ,\ \ldots  \ldots 
\eqno(4.2.7)$$
We now write down the linearised gauge transformations, that is, equation (3.11) in four  dimensions,  at low levels. These should     preserve the level and $SL(5)\otimes SL(8) $ character of the equations and so one finds that  
$$
\delta h_a{}^b= \partial_a \xi^b+ {1\over 2}  \delta _a^b \partial _{I_1I_2}\Lambda^{I_1I_2} + {1\over 2} \delta _a^b \partial ^{I_1I_2}\Lambda_{I_1I_2} + \ldots 
$$
$$
\delta \phi_{I_1\ldots I_4}= -6 \partial_{[ I_1I_2} \Lambda_{I_3 I_4] }
-{1\over 4} \epsilon_{I_1\ldots I_4 K_1\ldots K_4}\partial^{K_1K_2}\Lambda^{K_3K_4}+\ldots 
$$
$$
\delta \phi^I{}_J= 2\partial^{IL}\Lambda_{LJ} -2\partial_{JL}\Lambda^{LI} 
-{1\over 4} \delta _J^I(\partial^{KL}\Lambda_{LK} -2\partial_{KL}\Lambda^{LK} ) +\ldots 
$$
$$
\delta A_{aI_1I_2}= -\partial_a \Lambda_{I_1I_2} +  6\partial^{J_1J_2}\Lambda_{aJ_1J_2 I_1I_2} -4\partial_{[I_1| L}\Lambda_{a}{}^L{}_{|I_2]}- {1\over 2} \partial_{I_1I_2} \hat \xi _a+\dots 
$$
$$
\delta A^{I_1I_2}_a= - \partial_a \Lambda^{I_1I_2} -{1\over 4}   \epsilon ^{I_1I_2K_1K_2J_1\ldots J_4 } \partial_{K_1K_2}\Lambda_{a}{}_{J_1\ldots J_4}+4\partial^{[I_1| L}\Lambda_{a}{}^{|I_2]}{}_L+ {1\over 2} \partial^{I_1I_2} \hat \xi _a+\dots 
$$
$$
\delta A_{a_1a_2}{}^J{}_K= 4\partial_{[a_1} \Lambda _{a_2]}{}^J{}_{K}+\ldots  ,\quad \delta \hat h_{ab}= -2 \partial_{(a}\hat \xi_{b)}+\ldots 
$$
$$
\delta A_{a_1a_2}{}_{I_1\ldots I_4}= 12\partial_{[a_1} \Lambda_{a_2]I_1\ldots I_4} +\ldots 
\eqno(4.2.8)$$\
We should  have written the equations with an arbitrary constants in front of each term, but we have taken the liberty of inserting the numerical values that result from demanding that the left and right-hand sides of the equation transform as the adjoint representation of $E_{11}$. This is the calculation we now carry out by first computing the required rigid $E_{11}$ transformations of the objects that appear in the equation at the linearised level. 
\par
The rigid $E_{11}$ transformation generated by $g_0= e^{a^{aI_1I_2}R_{aI_1I_2}+a^{a}{}_{I_1I_2}R_{a}{}^{I_1I_2}} $, using equation (3.4),  leads to the transformations 
$$
\delta x^a= 2a^{aI_1I_2}x_{I_1I_2}-2a^{a}{}_{I_1I_2}x_{}^{I_1I_2},\ 
\delta x_{I_1I_2}= a^{a}{}_{I_1I_2}\hat x_a -12 a^{aJ_1J_2}x_{aJ_1J_2I_1I_2}
- 8 a^a {}_{K[I_1|} x_a{}^K{}_{|I_2]}
$$
$$
\delta x^{I_1I_2}= a^{a}{}^{I_1I_2}\hat x_a 
-{1\over 2}  a^{a}{}_{J_1J_2}x_{aJ_3\ldots J_6}\epsilon ^{J_1\ldots J_6 I_1I_2} 
- 8 a^a {}^{K[I_1|} x_a{}^{|I_2]}{}_{K}, \ldots 
\eqno(4.2.9)$$
The corresponding transformations of the generalised derivatives are given by 
$$
\delta (\partial_a )=0,\ \delta (\partial^{I_1I_2} )= -2 a^{a}{}^{I_1I_2} \partial_a,\ \delta (\partial_{I_1I_2} )= 2 a^{a}{}_{I_1I_2} \partial_a,
$$
$$
 \delta( \hat \partial _a )= -  a^{a}{}_{[I_1I_2}\partial ^{I_3I_4]}-
a^{a[I_1I_2}\partial _{I_3I_4]}, 
$$
$$
\delta (\partial^{aI_1\ldots I_4})= 12 a^{a[I_1I_2}\partial ^{I_3I_4]}
+{1\over 2} \epsilon ^{I_1\ldots I_4 J_1J_2 K_1K_2}a^a{}_{J_1J_2} \partial_{K_1K_2},\ 
$$
$$
\delta (\partial^a{}_K{}^J)=8 a^a{}_{KL} \partial^{LJ}
+8 a^a{}^{JL} \partial_{LK} , \ldots 
\eqno(4.2.10)$$
\par
We also need the variations of the fields at the linearised level. These 
follow from equation (3.3) and, after some work which includes compensating the group element using a  local $I_c(E_{11})$ transformation, one finds the result 
$$
\delta h_a{}^b= 2 A_a{}^{I_1I_2} a^b{}_{I_1I_2} -2  A_a{}_{I_1I_2} a^b{}^{I_1I_2} - \delta_a^b A_c{}^{I_1I_2} a^c{}_{I_1I_2} + \delta_a^b  A_c{}_{I_1I_2} a^c{}^{I_1I_2} 
$$
$$
\delta \phi_{I_1\ldots I_4}= 12 a^{a}{}_{[I_1I_2|}A_{a|I_3I_4]}
-{1\over 2} \epsilon_{I_1\ldots I_4 J_1\ldots J_4} a^{aJ_1J_2}A_a{}^{J_3J_4}
$$
$$
\delta \phi^J{}_K= -4A_{a}{}^{LJ} a^a{}_{LK} -4 A_a{}_{LK}a^{aLJ}
$$
$$
\delta A_a{}^{I_1I_2} =-4 a^b{}^{k[I_1|}A_{ba}{}^{|I_2]}{}_K
-{1\over 12} \epsilon ^{I_1I_2 J_1\ldots J_4 K_1K_2} a^b{}_{K_1K_2} A_{ab}{}_{J_1\ldots J_4} + a^{bI_1I_2}\hat h _{ab}
$$
$$+ (a^b{}_{J_1J_2} \hat h_{b}{}^{c}
+2a^c{}_{J_1J_2} -a^{cK_1K_2}\phi_{K_1K_2 J_1J_2}+2a^{c}{}_{LJ_2} \phi^{L}{}_{J_1})\eta_{ca}\delta ^{I_1I_2, J_1J_2}+\ldots
$$\
$$
\delta A_a{}_{I_1I_2} = +4 a^b{}_{J[I_1|}A_{ba}{}^{J} {}_{I_2]}
-2a^b{}^{K_1K_2} A_{ba}{}_{K_1K_2 I_1I_2} +a^b{}_{I_1I_2} \hat h_{ba}
$$
$$+(a^b{}^{J_1J_2} \hat h_{b}{}^{c}
+2a^c{}^{J_1J_2} -{1\over 4!} \epsilon ^{K_1\ldots K_4 L_1L_2 J_1J_2}
\phi^{c}_{L_1L_2} +2 \phi^{J_1}{}_{L} a^{c LJ_2}
)\eta_{ca}\delta _{I_1I_2, J_1J_2}+\ldots 
\eqno(4.2.11)$$
The last terms in $\delta A_a{}_{I_1I_2} $ and $\delta A_a{}^{I_1I_2}$ arise from the need, mentioned above, to  bring the group element into the required form using a local $I_c(E_{11})$ transformation. 
\par
To check that right-hand side of equation (4.2.8) belongs to the adjoint representations of $E_{11}$ we use equations (4.2.9) and  (4.2.10),  which give the rigid transformations of the gauge parameters and generalised derivatives respectively, to find the transformations of the right-hand side of equation (4.2.8). We then compare this with rigid transformations 
 of the fields,  given in equation (4.2.11), and so find the coefficients as given in equation (4.2.8).  
\par
To find the variations of the fields we need the generalised vielbein which is straightforward to calculate  using its definition in equation (3.8); one finds that [21] 

$$
{ E}= (det e)^{-{1\over 2}}
\left(\matrix {e_\mu{}^a&-
e_\mu{}^c A_{c}{}_{J_1J_2}& - e_\mu{}^c
A_{c}{}^{J_1J_2}\cr  0&{\cal N}^{I_1I_2}{}_{J_1J_2}& {\cal
N}^{I_1I_2}{}^{J_1J_2}
 \cr 0&{\cal N}_{I_1I_2}{}_{J_1J_2}& {\cal
N}_{I_1I_2}{}^{J_1J_2}
\cr}\right)
\eqno(4.2.12)$$
The matrix ${\cal N}$ is the vielbein in the scalar sector, which is given by 
$g_\phi^{-1}  (dx_{I_1I_2}Z^{I_1I_2}+   dx^{I_1I_2}Z_{I_1I_2})g_\phi
\equiv dx\cdot {\cal N} \cdot l $ where   $g_\phi $ is the group element for the non-linear realisation of $E_7$ with local subgroup SU(8); it just depends on the scalar fields.  
\par
We now compute the variation of equation (3.13) for the case the $A=a$ and $B=I_1I_2$. The left hand-side is given by 
$$
(E^{-1})_a{}^\Pi \delta E_\Pi{}_{I_1I_2}= -(e^{-1})_a{}^\mu \delta A_\mu{}_{\dot J_1\dot J_2}{\cal N} ^{\dot J_1\dot J_2}{}_{I_1I_2}
\eqno(4.2.13)$$
where $\dot I, \dot J,\ldots $ are curved indices. Evaluating the formula we find  several   terms, but the net effect is that the formula  contains   the variation of $A_{\mu \dot I_1\dot I_2}$, that is the vector with world   indices. Examining equation (4.2.8) we find that the right-hand side of equation (3.13), for the indices above, takes the from 
$$
-(E^{-1})_a{}^\Pi D_\Pi\Lambda _{I_1I_2}+ 
6(E^{-1})^{J_1J_2} {}^\Pi D_\Pi \Lambda _{a J_1J_2I_1I_2} 
-4 (E^{-1})_{[I_1|L}{}^\Pi D_\Pi \Lambda_a{}^L{}_{|I_2]} 
$$
$$
-{1\over 2} (E^{-1})_{I_1I_2}{}^\Pi D_\Pi\hat \xi_a+\ldots 
\eqno(4.2.14)$$
\par
Equating equations (3.2.12) and (3.2.13),  multiplying by the inverse vielbein in space-time and the vielbein in the 56-dimensional internal space,  
we find that 
$$
 \delta A_\mu{}_{\dot I_1\dot I_2}= - \{\tilde  D_\mu \Lambda {}_{\dot I_1\dot I_2}
+ A_\mu {}^{\dot K_1\dot K_2}\tilde  D_{\dot K_1\dot K_2}\Lambda {}_{\dot I_1\dot I_2}+ A_\mu {}_{\dot K_1\dot K_2} \tilde D^{\dot K_1\dot K_2}\Lambda {}_{\dot I_1\dot I_2}\}
$$
$$
+6\{ \tilde D^{\dot J_1\dot J_2} \Lambda_{\mu \dot J_1\dot J_2\dot I_1\dot I_2} -\Omega ^{\dot J_1\dot J_2}{}_{, \mu [\dot J_1\dot J_2}\Lambda _{\dot I_1\dot I_2] }+ {1\over 4!} \epsilon_{\dot I_1\dot I_2 \dot J_1\dot J_2\dot K_1\dot K_2\dot L_1\dot L_2} \Omega ^{\dot J_1\dot J_2}{}_{, \mu}{}^{ \dot K_1\dot K_2}\Lambda ^{\dot L_1\dot L_2}\}
$$
$$
-4\{ \tilde D_{[\dot I_1 |\dot L} \Lambda _{\mu}{}^{\dot L} {}_{|\dot I_2]}
+\Omega _{[\dot I_1|\dot L}{}_{, \mu \dot K |\dot I_2 ]} \Lambda ^{\dot K\dot L } 
+\Omega _{[\dot I_1|\dot L}{}_{, \mu}{}^{\dot L \dot K | } \Lambda _{\dot K |\dot I_2 ] }\}
$$
$$
-{1\over 2} \{-\Omega _{\dot I_1\dot I_2}{}_{,\mu }{}^{\dot K_1\dot K_2}\Lambda_{\dot K_1\dot K_2}
+\Omega _{\dot I_1\dot I_2}{}_{,\mu }{}_{\dot K_1\dot K_2}\Lambda^{\dot K_1\dot K_2}\}+ \ldots 
\eqno(4.2.15)$$
where $\tilde D_\Pi$ is the covariant derivative with a connection that contains only the level zero parts and 
$\Omega_{\dot I_1\dot I_2, \mu \dot J_1\dot J_2} = \partial_{\dot I_1\dot I_2} A_{\mu \dot J_1\dot J_2}$.  
We note that although equation (3.13) is simple,  its explicit form for a given field can appear complicated. 
\par
As we have mentioned, the non-linear realisation of $E_{11}\otimes_s l_1$ 
has been constructed at low levels, at least for  the vector and scalar fields [21]. These equations include terms that have the generalised space-time derivatives 
$\partial_{I_1I_2}$ and $\partial_{I_1I_2}$ and we refer the reader to this reference for the details.  However, it is 
 instructive to reproduce  part of the vector  equation of  motion  using the rigid transformations given in this paper. The vector equation 
carries two anti-symmetrised Lorentz  indices and  belongs to the $28+\overline { 28}$ -dimensional representations of SL(8).  The lowest level terms in the $28$ part of this equation are of the form 
$$
E_{a_1a_2 I_1I_2}\equiv {\cal F}_{a_1a_2 I_1I_2}\pm {i\over 2} \epsilon _{a_1a_2 b_1b_2} {\cal F}^{b_1b_2 }{}_{I_1I_2}+\ldots =0 
\eqno(4.2.16)$$  
where 
$$
{\cal F}_{a_1a_2I_1I_2}\equiv  \partial_{[a_1} A_{a_2] I_1 I_2 } +d_1 \partial^{J_1J_2}
A_{a_1a_2 J_1J_2 I_1I_2} + d_2\partial_{[I_1| L }A_{a_1a_2}{}^L{}_{|I_2]} +\dots  
\eqno(4.2.17)$$
and  $d_1$ and $d_2$ are constants.
The variation of the vector equation under the transformation of equation 
(4.2.11) must give one of the other equations of motion and in particular the scalar equation of motion.  This is an equation that has a single  Lorentz index and involves the derivatives of the scalars and  the epsilon symbol acting on the  the dual scalar fields, for example 
$\epsilon^{ab_1b_2b_3} \partial_{b_1}A_{b_2b_3 J_1J_2 I_1I_2}$.
 However, this is only the case  if we take the coefficients $d_1=-{1\over 2}$ and $d_2 =1$ and then   the variation is given by 
$$
\delta ({\cal F}_{ a_1a_2 I_1I_2})= 3a^{bJ_1J_2} \partial _{[b} A_{a_1a_2 ]J_1J_2 I_1I_2} 
+ 6 a^b{}_{[I_1| L }\partial_{[b} A_{a_1a_2]}{}^L{}_{|I_2]}+\ldots 
\eqno(4.2.18)$$
Taking into account the change of notation this agrees with the results of reference [21] whose fully non-linear results were found using the more powerful method of working with the Cartan forms and their local transformations. 
\par
It is  straight forward to verify, at the linearised level and up  to the level considered,  that the equation of motion for the vector of equation (4.2.17) is invariant under the gauge transformation of equation (4.2.8)  if we take the constants to have the  values determined by the symmetries of the $E_{11}\otimes_s l_1$ non-linear realisation.

\par
We close this section by sketching the gauge variations of the fields at linearised level  when one includes the  level three gauge parameters given in equation (4.2.7). The transformations of the level two  fields under the level three gauge parameters must involve the level $-1$ generalised space-time derivatives and have the generic form 
$$
\delta A_{a_1a_2}{}_{I_1\ldots I_4} = \ldots + \partial_{[I_1I_2|} 
\Lambda _{[a_1 ,a_2]}{}_{|I_3 I_4]}+ \partial_{[I_1| J} \Lambda_{[a_1 ,a_2 ]}{}_{| I_2I_3I_4]}{}^{J}+\ldots 
$$
$$
+\epsilon _{I_1\ldots I_4 K_1\ldots K_4} (\partial^{K_1K_2} \Lambda _{[a_1 ,a_2]}{}^{K_3K_4} + 
\partial^{K_1J} \Lambda _{[a_1 ,a_2]}{}^{K_2K_3K_4}{}_{J})+\ldots , \quad 
$$
$$
\delta A_{a_1a_2}{}^{I}{}_{J}=\ldots +  \partial^{IK}\Lambda_{[a_1 ,a_2]}{}_{KJ}
+\partial_{JK}\Lambda_{[a_1 ,a_2]}{}^{IK}
+\partial^{K_1K_2}\Lambda_{[a_1 ,a_2]}{}_{K_1K_2 J}{}^{I} 
+\partial_{K_1K_2}\Lambda_{[a_1 ,a_2]}{}^{K_1K_2 I}{}_{J} +\ldots 
$$
$$
\delta \hat h_{ab} = \ldots + \partial^{I_1I_2} \Lambda _{(a_1, a_2)}{}_{I_1I_2} 
+\partial_{I_1I_2} \Lambda _{(a_1, a_2)}{}^{I_1I_2} +\ldots 
\eqno(4.2.18)$$
While the lowest level transformations of the level three fields are given by 
$$
\delta A_{a_1a_2,b }= \partial_b\Lambda_{[a_1,a_2]} {}^{I_1I_2} + 
\partial_{[a_1}\Lambda_{a_2], b}{}^{I_1I_2}+\ldots 
$$
$$
\delta A_{a_1a_2,b }= \partial_b\Lambda_{[a_1,a_2]} {}_{I_1I_2} + 
\partial_{[a_1}\Lambda_{a_2], b}{}_{I_1I_2}+\ldots 
$$
$$
\delta A_{a_1a_2a_3 }{}_{\bullet} = \partial_{ [ a_1}\Lambda_{a_2a_3]}{}_{\bullet} +\ldots 
\eqno(4.2.19)$$
where $\bullet $  stands for the indices of the 912 representation of $E_7$ . 
It is straightforward to find the coefficients in front of the above terms using the methods used at lower levels.


\medskip 
{\bf 4.3 Gauge transformations in eleven   dimensions}
\medskip

We now consider the eleven  dimensional theory which emerges by decomposing $E_{11}$, and the $l_1$ representation,  with respect to its  SL(11) subalgebra.  In this decomposition the positive level generators of $E_{11}$  are  given by [1] 
$$
K^a{}_b (0) ;\  R^{a_1a_2a_3}(1);\  R^{a_1\ldots a_6} (2) ;\ 
R^{a_1\ldots a_8,b} (3) , \ldots 
\eqno(4.3.1)$$ 
where the   figures in the brackets indicate the level. 
The corresponding fields in the $E_{11}\otimes_s l_1$ non-linear realisation are  given by 
$$
h_a{}^b ; \ A_{a_1 a_2a_3};\  A_{a_1\ldots a_6} ;\ 
h_{a_1\ldots a_8,b}, \ldots 
\eqno(4.3.2)$$
Decomposing  the $l_1$ representation into representations of $SL(11)$ we find that [2]
$$
P_a (0),\  Z^{a_1a_2}  (1), ,\  Z^{a_1\ldots a_5}  (2), 
,\  Z^{a_1\ldots a_8}  (3),\ ,\  Z^{a_1\ldots a_7,b}  (3), 
$$
$$
Z^{b_1b_2b_3, a_1\ldots a_8} (4) , \quad
Z^{(b_1b_2) , a_1\ldots a_9} (4), \quad 
Z^{b_1b_2 , a_1\ldots a_9} (4) , \quad 
Z^{b , a_1\ldots a_{10}} (4) , \quad Z (4), 
\dots 
\eqno(4.3.3)$$
The second to last generator in the last line has multiplicity two meaning that the $l_1$ representation  contains two copies of it. In the non-linear realisation these lead to a generalised space-time  which  has  the  coordinates 
$$
x^a (0),\ x_{a_1a_2}  (1), ,\  x_{a_1\ldots a_5}  (2), 
,\  x_{a_1\ldots a_8}  (3),\ ,\  x_{a_1\ldots a_7,b}  (3), 
$$
$$
x_{b_1b_2b_3, a_1\ldots a_8} (4) , \quad
X_{(b_1b_2) , a_1\ldots a_9} (4), \quad 
x_{b_1b_2 , a_1\ldots a_9} (4) , \quad 
X_{b , a_1\ldots a_{10}} (4) , \quad x (4), 
\dots 
\eqno(4.3.4)$$
We denote the corresponding generalised derivatives by 
$$
\partial_a (0),\ \partial^{a_1a_2}  (-1), ,\  \partial^{a_1\ldots a_5}  (-2), 
,\  \partial^{a_1\ldots a_8}  (-3),\ ,\  \partial^{a_1\ldots a_7,b}  (-3), 
\dots  , 
$$
$$
\partial^{b_1b_2b_3, a_1\ldots a_8} (-4) , \quad
\partial^{(b_1b_2) , a_1\ldots a_9} (-4), \quad 
\partial^{b_1b_2 , a_1\ldots a_9} (-4) , \quad 
\partial^{b , a_1\ldots a_{10}} (-4) , \quad \partial (-4), \ldots 
\eqno(4.3.5)$$
where the number in brackets is the level. They are given in terms of the coordinates in the way expected, for example, 
$\partial_a = {\partial\over \partial x^a},\  \partial^{a_1a_2} = {\partial\over \partial x_{a_1a_2}}$. 
\par
The non-linear realisation of $E_{11}\otimes_s l_1$ was systematically constructed at low levels in reference [20] for the above fields and  the coordinates at levels zero,  one and two. The equations of motion for the three form and six form,   when truncated to the usual fields and coordinates,  are indeed those of the eleven  dimensional maximal supergravity theory. The field equations for gravity involves the usual field of gravity and the dual gravity field and these are  the subject of further study [31]. Furthermore,  once one assumes that    these equations possess  the expected   $SL(11) $ character, then  the symmetries of the $E_{11}\otimes_s l_1$ non-linear realisation appear to determine these equations uniquely. 
\par
According to the  discussion of section three  we should introduce 
local symmetries with  parameters that are in one to one correspondence with the $l_1$ representation which is, by construction,   in one to one correspondence with the coordinates of equation (4.2.4). As a result  the local parameters  are given by 
$$
\xi^a (0),\  \Lambda_{a_1a_2}  (1), ,\  \Lambda_{a_1\ldots a_5}  (2), 
,\  \Lambda_{a_1\ldots a_8}  (3),\ ,\  \Lambda_{a_1\ldots a_7,b}  (3), 
$$
$$
\Lambda_{b_1b_2b_3, a_1\ldots a_8} (4) , \quad
\Lambda_{(b_1b_2) , a_1\ldots a_9} (4), \quad 
\lambda_{b_1b_2 , a_1\ldots a_9} (4) , \quad 
\Lambda _{b , a_1\ldots a_{10}} (4) , \quad \Lambda  (4), 
\dots 
\eqno(4.3.6)$$
We now write down the linearised gauge transformations  at low levels, that is find equation (3.11) in eleven   dimensions. These equations should    preserve the level and $SL(11) $ character; the result is as follows   
$$
\delta h_a{}^b= \partial_a \xi^b-2   \partial ^{bc}\Lambda_{a}{}_{c}
+{1\over 3} \delta_a^b  \partial ^{c_1c_2}\Lambda_{c_1c_2}+\ldots
$$
$$
\delta A_{a_1a_2a_3}= -\partial_{[a_1} \Lambda_{a_2a_3]}-10 \partial^{c_1c_2}\Lambda_{c_1c_2 a_1a_2a_3}-{1\over 6} \partial_{a_1a_2a_3}{}^{b_1b_2}\Lambda_{b_1b_2} -{1\over 2} \partial_{[a_1a_2}\xi_{a_3]}+\ldots ,
$$
$$ 
\delta A_{a_1\dots a_6}= 2\partial_{[a_1} \Lambda_{a_2\ldots a_6]}+\ldots
\eqno(4.3.7)$$
We have taken the liberty of inserted the values of the coefficients that follow from the analysis given below.  We note the appearance of the last two terms in the variation of the three form whose indices are not in the expected places. These arise from the need to compensate the group element by a local  $I_c(E_{11})$ transformation in order to preserve its form. 
\par
We now compute the transformations of the coordinates and fields under the rigid $E_{11}$ transformation of the form $g_0= e^{a^{a_1a_2a_3} R_{a_1a_2a_3}}$. Using equation (3.3) we find that the coordinates transforms as 
$$
\delta x^a=6x_{b_1b_2} a^{b_1b_2 a},\ \delta x_{a_1a_2}= {5!\over 2} x_{b_1b_2b_3a_1a_2a_3}a_{b_1b_2b_3},\ldots 
\eqno(4.3.8)$$
and, as a result,  the generalised derivatives transform as 
$$
\delta (\partial_a )=0,\ \delta (\partial ^{a_1a_2})= -6  a_{a_1a_2b}\partial_b,\ 
\delta (\partial^{a_1\dots a_5})= -{5!\over 2} a^{[a_1a_2a_3}\partial^{a_4a_5]},\ \dots 
\eqno(4.3.9)$$
\par
The linearised transformations of the fields under the same rigid $E_{11}$ transformation  is given by 
$$
\delta h_a{}^b= -18 a^{c_1c_2 b} A_{c_1c_2 a} + 2 \delta _a^b a^{c_1c_2 c_3} A_{c_1c_2 c_3} , 
$$
$$ 
\delta A_{c_1c_2 c_3} = -60 a^{c_1c_2 c_3}A_{c_1c_2 c_3a_1a_2a_3 }
+ (2 a^{b_1b_2b_3} +3 h^{[ b_1|}{}_d a^{d|b_2b_3 ]})\delta_{a_1a_2a_3, b_1b_2b_3}  
$$
$$
\delta A_{c_1\dots  c_6} =- 16.7 a^{c_1c_2 c_3}h_{c_1c_2 c_3a_a\ldots a_5,  a_6 }+16.7 a^{c_1c_2 c_3}h_{c_1c_2 a_1\ldots a_6,  c_3}+\ldots
\eqno(4.3.10)$$
\par
Using equations (4.3.8) and (4.3.9) to compute the transformation of the 
right-hand side of equation (4.3.7), and comparing this with the transformation of the left-hand side, given in equation (4.3.10), we find the coefficients given in equation (4.3.7). 
\par
Finally we find the non-linear gauge variation of the three form field. To this end we require the generalised vielbein defined in equation (3.8) which is given at low levels by [20]
$$
{ E}= (det e)^{-{1\over 2}}
\left(\matrix {e_\mu{}^a&-3 e_\mu{}^c A_{cb_1b_2}& 3 e_\mu{}^c A_{cb_1\ldots b_5}+{3\over 2} e_\mu{}^c A_{[b_1b_2b_3}A_{|c|b_4b_5]}\cr
0&(e^{-1})_{[b_1}{}^{\mu_1} (e^{-1})_{b_2]}{}^{\mu_2}&- A_{[b_1b_2b_3 }(e^{-1})_{b_4}{}^{\mu_1} (e^{-1})_{b_5 ]}{}^{\mu_2}  \cr
0&0& e^{-1})_{[b_1}{}^{\mu_1} \ldots (e^{-1})_{b_5]}{}^{\mu_5}\cr}\right)
\eqno(4.3.11)$$
We take the indices to take the values $A=a_1$ and $B=a_2a_3$ and for these the left-hand side of equation (3.13) is given by 
$$
(E^{-1})_A{}^\Pi \delta E_\Pi{}^B= 
(e^{-1})_{a_1}{}^{\mu_1}(e^{-1})_{a_2}{}^{\mu_2}(e^{-1})_{a_3}{}^{\mu_3}
\delta A_{\mu_1\mu_2\mu_3}
\eqno(4.3.12)$$
It contains the variation of the three form gauge field with world indices. 
Looking at equation (4.3.7) we find that the right-hand side is given by 
$$
-(E^{-1})_{[a_1|}{}^\Pi D_\Pi \Lambda_{|a_2a_3]}- 10(E^{-1})^{b_1b_2}{}^\Pi D_\Pi \Lambda_{b_1b_2 a_1a_2a_3}+\ldots 
$$
$$
= -(e^{-1})_{[a_1|}{}^\mu \tilde D_\mu \Lambda_{|a_2a_3]}-10 (E^{-1})^{b_1b_2}{}_{\nu_1\nu_2} \tilde D^{\nu_1\nu_2} \Lambda_{b_1b_2 a_1a_2a_3}
$$
$$
= - (e^{-1})_{[a_1|}{}^\mu \tilde D_\mu \Lambda_{|a_2a_3]}-10  e_{\nu_1}{}^{b_1} e_{\nu_2}{}^{b_2} (\tilde  D^{\nu_1\nu_2}\Lambda_{b_1b_2 a_1a_2a_3}- \Omega^{\nu_1\nu_2} {}_{, a_1a_2a_3} \Lambda_{c_1c_2 })+ \ldots 
\eqno(4.3.13)$$
where $\tilde D_\mu$ is the covariant derivative with only the connections for Lorentz transformations. Equating these one can  read off the gauge transformation of the three form and one finds it to be given by 
$$ 
\delta A_{\mu_1\mu_2\mu_3}= -\partial_{[\mu_1}\Lambda_{\mu_2\mu_3]} 
-3 A_{[\mu_1 |\nu_1\nu_2} \tilde D^{\nu_1\nu_2}\Lambda_{|\mu_2\mu_3]}
-10 (\tilde D^{\nu_1\nu_2} \Lambda_{\nu_1\nu_2 \mu_1\mu_2\mu_3} 
+ \Omega^{\nu_1\nu_2}{}_{, \mu_1\mu_2\mu_3} \Lambda_{\nu_1\nu_2})+\ldots 
\eqno(4.3.14)$$


\medskip 
{\bf 5. The  Explicit form of the Gauge transformations }
\medskip
In this section we will further develop the general theory underlying the gauge transformations and  give  explicit formulae for the gauge transformation for all fields in the non-linear realisation.    We begin by 
considering  the generalised spin connection in more detail. The tangent algebra of the generalised tangent space-time is the Cartan involution invariant subalgebra 
of $E_{11}$,  denoted $I_c (E_{11})$. We can introduce a corresponding generalised spin connection which is valued in $I_c (E_{11})$; 
$$
\omega_\Pi{}\equiv \omega_{\Pi ,}{}_{\underline \alpha} S^{\underline \alpha}
\eqno(5.1)$$
where $S^{\underline \alpha}$ are the generators of  $I_c (E_{11})$. The generalised spin connection transforms as 
$$
\omega_\Pi{}^\prime = h^{-1}\omega_\Pi{}h + h^{-1} \partial_\Pi h 
\eqno(5.2)$$
This is just like the parts of the Cartan form that belong to the subalgebra 
$I_c(E_{11})$ but as is clear from the lowest level component, that is, gravity, the generalised spin connection is not in general simply  equal to these  components of the 
Cartan form, see the expression below equation (2.3). 
The associated  covariant derivative is given by 
$$
D_\Pi= \partial _\Pi + \omega_\Pi
\eqno(5.3)$$
and the corresponding generalised curvatures are defined  by 
$$
[D_\Pi , D_\Lambda ]= R_{\Pi\Lambda }\equiv R_{\Pi\Lambda , }
{}_{\underline \alpha} S^{\underline \alpha}
\eqno(5.4)$$
We can also construct the analogue of the generalised torsions 
$$
T_{\Pi\Lambda }\equiv  T_{\Pi\Lambda ,}{}^A l_A= \partial_\Pi E_\Lambda  +[ \omega _\Pi , E_\Lambda ]
-(\Pi \leftrightarrow \Lambda)
\eqno(5.5)$$
where $E_\Pi= E_\Pi{}^A l_A$. The generalised curvature transforms under local $I_c(E_{11})$ transformations as 
$R_{\Pi\Lambda }{}^\prime= h^{-1}R_{\Pi\Lambda }h$ and the torsion in the same way as the vielbein on its upper index. 
\par
To gain a better idea of what these objects contain we will compute the above quantities in {\bf eleven dimensions}. The generalised spin connection  is given by 
$$
\omega_\Pi{}= {1\over 2} \omega _{\Pi ,}{}^a{}_b J^a{}_b 
+ \omega _{\Pi ,}{}_{a_1a_2a_3} S^{a_1a_2a_3}+ 
+ \omega _{\Pi ,}{}_{a_1\ldots a_6} S^{a_1\ldots a_6}+ \ldots 
\eqno(5.6)$$
where the generators of $I_c (E_{11})$ and their commutators can be found in references [2],  or [5], for example. The curvatures are simple to calculate and are given by 
$$
R_{\Pi\Lambda , }{}^a{}_{b} = 
\partial_\Pi \omega _{\Lambda ,} {}^a{}_b +\omega _{\Pi ,}{}^a{}_c  \omega _{\Lambda , }{}^c{}_b  
-18 \omega _{\Pi ,}{}^{c_1c_2 a}\omega _{\Lambda , }{}_{c_1c_2 b}+\ldots 
-(\Pi \leftrightarrow \Lambda), \quad 
$$
$$
R_{\Pi\Lambda ,}{}_{a_1a_2a_3} =  \partial_\Pi \omega _{\Lambda , } {}^{a_1a_2a_3} +3\omega _{\Pi ,}{}^{[ a_1|}{}_c  \omega _{\Lambda , }{}^{ c |a_2a_3 ]}  
-{5!\over 2}  \omega _{\Pi ,}{}_{c_1c_2 c_3}\omega _{\Lambda ,}{}^{c_1c_2 c_3 a_1a_2a_3}
+\ldots
-(\Pi \leftrightarrow \Lambda)
 , \quad 
$$
$$
R_{\Pi\Lambda ,}{}_{a_1\ldots a_6} =  \partial_\Pi \omega _{\Lambda , } {}^{a_1\ldots a_6} +6\omega _{\Pi ,}{}^{[ a_1|}{}_c  \omega _{\Lambda , }{}^{ c |\ldots a_6  ]}  
+   2\omega _{\Pi , }{}^{[ a_1a_2 a_3}\omega _{\Lambda , }{}^{a_4a_5a_6 ]}  +\ldots
-(\Pi \leftrightarrow \Lambda), \ldots  
\eqno(5.7)$$
while  the torsions are given by 
$$
T_{\Pi\Lambda , }{}^b= \partial_\Pi E_\Lambda {}^{b} +\omega _{\Pi ,}{}^b{}_c E_\Lambda {}^{c} 
-6\omega _{\Pi ,}{}^{bc_1c_2}  E_{\Lambda ,} {}_{c_1c_2} 
 +3.5!\omega _{\Pi ,}{}^{bc_1\ldots c_5}  E_{\Lambda ,} {}_{c_1\ldots c_5} +\ldots -(\Pi \leftrightarrow \Lambda), 
$$
$$
T_{\Pi\Lambda , }{}_{a_1a_2} = \partial_\Pi E_{\Lambda ,}  {}_{a_1a_2} 
+3\omega _{\Pi ,}{}^{da_1a_2}  E_\Lambda {}_{d} 
-2\omega _{\Pi ,}{}^{b}{}_{[a_1 |}  E_{\Lambda ,} {}_{|a_2 ] b} 
$$
$$-{5!\over 2} \omega _{\Pi ,}{}^{c_1c_2c_3}  E_{\Lambda ,}  {}_{c_1c_2c_3 a_1a_2}  +\ldots 
-(\Pi \leftrightarrow \Lambda),
$$
$$
T_{\Pi\Lambda ,}{}_{a_1\ldots a_5} = \partial_\Pi E_{\Lambda ,}  {}_{a_1\ldots a_5} 
-3\omega _{\Pi ,}{}^{da_1\ldots a_5}  E_\Lambda {}_{d} 
$$
$$
+5 \omega _{\Pi ,}{}_{[a_1|}{}^{b}  E_{\Lambda ,}  {}_{b|a_2\ldots a_5 ]} 
+\omega _{\Pi ,}{}_{[a_1 a_2a_3}  E_{\Lambda ,} {}_{a_4a_5 ]}+ \ldots 
-(\Pi \leftrightarrow \Lambda), \ldots 
\eqno(5.8)$$
\par
We have for clarity introduced in this equation a comma to separate the world and tangent indices on the generalised vielbein. We note that the construction of the Cartan involution invariant subalgebra 
$I_c(E_{11})$ does not preserve the level and as a result neither do the generalised curvatures and torsions. It is very straightforward to compute the analogous quantities in five and four dimensions, or indeed in any other dimension. 
\par
It would be natural to take the spin connection to satisfy the equation 
$$
T_{\Pi\Lambda ,}{}^A=0 
\eqno(5.9)$$
At the lowest level this is just the equation that solves the spin connection of general relativity in terms of the vielbein.  
\par
Before considering the gauge transformations it will be instructive to consider the construction of the generalised vielbein in more detail. We can write the $E_{11}\otimes_s l_1$ algebra in the form 
$$
[R^{\underline \alpha} , R^{\underline \beta} ]= f^{\underline \alpha \underline \beta}{}_{\underline \gamma} R^{\underline \gamma}, \quad
[R^{\underline \alpha} , l_A]= -(D^{\underline \alpha} )_A{}^B l_B
\eqno(5.10)$$
The Jacobi identities imply that the matrices $(D^{\underline \alpha} )_A{}^B$ are a representation of the $E_{11}$ algebra, that is, 
$[D^{\underline \alpha}, D^{\underline \beta }]= f^{\underline \alpha \underline \beta}{}_{\underline \gamma} D^{\underline \gamma}$. These matrices are related to the matrices of the $l_1$ representation given in section three, namely $D(I+a_{\underline \beta } R^{\underline \beta })= 
I+ a_{\underline \beta } D^{\underline \beta }$ to lowest order in 
$a_{\underline \beta }$. 
\par
The vielbein is defined in equation (3.8) and if we take the $E_{11}$ group element to be of the form $g_E= e^{A_{\underline \alpha }R^{\underline \alpha }}$ then the generalised vielbein is given by 
$$
E_\Pi{}^A= (e^{ {\cal A}})_\Pi {}^A
\eqno(5.11)$$
where the matrix ${\cal A}$ is given by 
$({\cal A})_\Pi {}^A= A_{\underline \alpha }( D^{\underline \alpha })_\Pi {}^A $ and one evaluates equation (5.11) by expanding the exponential in the usual way and taking the product to be matrix multiplication. We note that we can write the fields either as $A_{\underline \alpha} $, or 
${\cal A}$, which are related as in the previous sentence. 
\par
In section three we discussed the general criterion that the gauge transformations must satisfy and in section four we used this to derive the gauge transformation in four, five and eleven dimensions for certain fields. Clearly one can carry this out for all fields as a matter of principle but we now give an  explicit expression for the gauge transformations. 
We begin by considering the linearised theory and take the variation of the gauge fields to be given by 
$$
\delta A_{\underline \alpha} = (D _{\underline \alpha})_\Pi{}^\Lambda\partial_\Lambda \Lambda ^\Pi
\eqno(5.12)$$
In other words we take the constants in equation (3.11) to be given by 
$N _{\underline \alpha}{}_\Pi{}^\Lambda= (D _{\underline \alpha})_\Pi{}^\Lambda$. Carrying out a  rigid $g_0\in E_{11}$ transformations on the parameter, $\Lambda ^{\Pi\prime}= \Lambda ^\Xi  D(g_0^{-1}) _\Xi{}^\Lambda$, and on the generalised derivative,  $\partial^\prime_ \Lambda = D(g_0) _\Lambda {}^\Delta\partial_\Delta  $, 
  we find that the right-hand side of equation (5.12) transforms as 
$$
(D _{\underline \alpha})_\Pi{}^\Lambda \partial_\Lambda^\prime  \Lambda ^{\Pi\prime} = D(g_0^{-1} R_{\underline \alpha } g_0 )
_\Pi{}^\Lambda \partial_\Lambda  \Lambda ^{\Pi} = 
(D _{\underline \alpha})_\Pi{}^\Lambda \partial_\Lambda \Lambda ^{\Pi }+ a_{\underline \beta} f^{\underline \beta \underline \gamma}{}_{\underline \alpha} (D _{\underline \gamma})_\Pi{}^\Lambda \partial_\Lambda \Lambda ^{\Pi }
\eqno(5.13))$$
which is indeed the appropriate $E_{11}$ transformation of $\delta A_{\underline \alpha} $. In deriving this result we have used that 
 $D$ is a representation of $E_{11}$ and we have taken $g_0= I+ a_{\underline \beta}  R^{\underline \beta}$, where  $a_{\underline \beta}$ are constants and worked to lowest order in this constant.   
We raise and lower indices with the Cartan-Killing metric $g_{\underline \alpha ,\underline \beta}$ which  vanishes unless $\underline \alpha +\underline \beta=0$.  As a result $D_{\underline \alpha}  =  g_{\underline \alpha , \underline \beta}
D^{\underline \beta}  
= g_{\underline \alpha , - \underline \alpha}
D^{-\underline \alpha}$
\par
We can express the variation of the fields of equation (5.12) in a more 
formal way. As the gauge parameters $\Lambda $ belong to  the $l_1$ representation we can write them in the form   
$$
\Lambda = \Lambda ^A l_A
\eqno(5.14)$$
We note that at the linearised level tangent and world indices are equivalent. As we have mentioned,  the generalised derivatives can be thought of as belonging to the representation $\bar l_1$. This has the generators $\bar l^A $ which can be taken to be the Cartan involution of those in $l_1$, that is, $I_c(l_A)= -\bar l^A$. As such we can package  the generalised derivatives in the object $\partial$ where  
$$
\partial= \partial_A \bar l^A
\eqno(5.15)$$
We can then write equation (5.12) as 
$$
\delta A_{\underline \alpha} = [R_{\underline \alpha} , \partial ]_A \Lambda ^A= ([R_{\underline \alpha} , \partial ] ,  \Lambda )
\eqno(5.16)$$
In the last line of this equation we have used the $E_{11}$ invariant scalar product between elements of the $l_1$ representation and those of the $l_1$ representation denoted $(\bar l , l)$. This formula provides a much quicker way to compute the linearised gauge variations as $[R_{\underline \alpha} , \partial ] $ is straight forward to compute using the known commutators between the $E_{11}$ generators and those of the $l_1$ representations and the action of the Cartan involution. The one point that requires a bit more work is to compute the invariant scalar product between the $l_1$ and $\bar l_1$ representations. The general method for determining this scalar product is explained in appendix A of reference [32] and it is found at low levels in eleven dimensions. 
\par
As we have mentioned the fields can also be encoded in  the matrix ${\cal A}_A{}^B \equiv A_{\underline \alpha} (D_{\underline \alpha } )_A{}^B$ and then the linearised variation of equation (5.12) can be written as 
$$
\delta {\cal A} _A{}^B = (D_{\underline \alpha } )_A{}^B
(D_{\underline \alpha } )_C{}^D \partial_D \Lambda^C
\eqno(5.17)$$
We note that at lowest order in the fields,  $E_\Pi{}^A= {\cal A} _\Pi{}^A$  and so also to lowest order $(E^{-1})_A{}^\Pi \delta E_\Pi{}^B= \delta {\cal A} _A{}^B$. As a result equation (5.17) has an obvious non-linear generalisation and we take the non-linear gauge transformations to be given by 
$$
E^{-1}{}_A{}^\Pi \delta E_\Pi {}^B = (D_{\underline \alpha } )_A{}^B 
(D_{\underline \alpha } )_C{}^D D_D \Lambda^C
\eqno(5.18)$$
where as before $\Lambda ^A= \Lambda^\Pi E_\Pi{}^A$ and 
$D_A\equiv  E^{-1}{}_A{}^\Pi D_\Pi = E^{-1}{}_A{}^\Pi ( \partial_\Pi +\Omega _\Pi ) $. If one uses the local subgroup $I_c(E_{11})$ to choose the group element to be in the Borel subgroup  one will have to carry out an local $I_c(E_{11})$ transformation at the same time as that of equation (5.18) to preserve the form of the vielbein. 
\par
The reader may find it consoling to evaluate this gauge transformation at lowest level, that is for gravity. In this case $(D^a{}_b)_c{}^d=\delta ^a_c\delta^d_b$ and substituting this in we find the general coordinate transformation of equation (2.3). 
\par
Under a rigid $g_0\in E_{11}\otimes_s l_1$ transformation of the form $g_0=e^{a^Al_A + a_{\underline \beta}R^{\underline \beta}}$ the generalised coordinates change as $z^A{}^\prime = z^A+a^A- z^B a_{\underline \beta} (D^{\underline \beta}) _B{}^A\equiv z^A+\tilde\Lambda^A$. As such we can treat these rigid $E_{11}\otimes_s l_1$ transformations as particular gauge transformation with parameter $\tilde\Lambda^A$. Substituting this gauge transformation in equation (5.18) one does indeed recover the known action of this rigid transformation on the generalised vielbein and so the fields $A_{\underline \alpha}$ up to and including order $O(A_{\underline \alpha})$. This calculation is a little more non-trivial than it appears at first sight and so we will not give it here. 
\par
We now consider the closure of the non-linear gauge transformations of equation (5.18) and local 
$I_c(E_{11})$ transformations. Let us consider a gauge transformation followed by a finite local $h\in I_c(E_{11})$ transformation on the vielbein; 
$$
E_\Pi{}^A{}\to E_\Pi{}^A{}^\prime = (D_\alpha)_B{}^A E_\Pi{}^E D(h) _E{}^B 
(D_\alpha)_C{}^D D(h^{-1})_D{}^F D(h) _G{}^C D_F \Lambda^G
$$
$$
= D(h)_F{}^A (D(h R_\alpha h^{-1})_B{}^F- \delta_B{}^F) E_\Pi{}^E 
(D(hR_\alpha h^{-1})_C{}^D-\delta _C{}^D)  D_D \Lambda^C= D(h)_F{}^A \delta_\Lambda E_\Pi {}^F
\eqno(5.19)$$
Taking an infinitesimal  local $ I_c(E_{11})$ transformation we find that it commutes with the gauge transformations,  as it did for gravity. We have used that the sum over the roots $\alpha$ is just rearranged by conjugation with $h$ and that $\Omega_\Pi$ behaves like a generalised spin connection.  
\par
We now find the variation of the Cartan forms under the gauge transformations. 
The Cartan forms ${\cal V}_E$ belong to the $E_{11}$ Lie algebra 
and taking the generators of $E_{11}$ to be in the first fundamental representation, that is the $l_1$ representation,  they can be written as $G_\Pi{}_A{}^B\equiv G_\Pi{}_{\underline\alpha} (D^{\underline\alpha})_A{}^B$.  
As we explained above,  the generalised vielbein can be thought of as being the transformation $g_E$ in the $l_1$ representation,  
that is, $E=D(g_E)$ and as a result we can write 
$$
G_{\Pi ,}{}_B{}^C = (E^{-1})_B{}^\Lambda \partial _\Pi E_\Lambda{}^C ,\quad  {\rm or} \quad 
G_{A,}{}_B{}^C \equiv (E^{-1})_A{}^\Pi G_\Pi{}_B{}^C 
= (E^{-1})_A{}^\Pi(E^{-1})_B{}^\Lambda \partial _\Pi E_\Lambda{}^C 
\eqno(5.20)$$
Writing the Cartan form in this way makes it easy to find an expression for its gauge variation and it is then straight forward to show that  
$$
\delta G_{A,}{}_B{}^C = {\cal N}_A{}^D G_{D,} {}_B{}^C- {\cal N}_B{}^D G_{A,}{}_D{}^C + G_{A,}{}_B{}^D  {\cal N}_D{}^B 
+ (E^{-1})_B{}^\Lambda\partial_\Lambda {\cal N}_A{}^B
\equiv {\cal D}_A {\cal N}_B{}^C
\eqno(5.21)$$
where  ${\cal N}_A{}^B\equiv  (E^{-1})_A{}^\Pi\delta  E_\Pi{}^B$.

\medskip 
{\bf 6.  Conclusions}
\medskip
In this paper we have proposed  gauge transformations for  all the fields in the $E_{11}\otimes_s l_1$ non-linear realisation. The gauge parameters are in one to one correspondence with the first fundamental representation of $E_{11}$, the $l_1$ representation and so can be thought of as generalised diffeomorphisms of the generalised space-time. The gauge transformations of even the familiar fields, for example the form fields,  arise from coordinates in the $l_1$ representation that are at a higher level than the usual space-time and even than the level one coordinates that are Lorentz scalars  (see the first column of the table). The $E_{11}$ fields which are beyond those usually used to formulate supergravity theories  generally  have  with mixed symmetry indices  and it is to be  expected that they will  have  gauge transformations with several different parameters.  This feature is  correctly accounted for by the gauge transformation proposed in this paper in that these fields are associated with more than one coordinate in the $l_1$ representation and so  do possess more than one gauge parameter. In section three we gave the general criterion that the gauge transformations must satisfy for them to be compatible with $E_{11}$ and this lead us to the gauge variation given in equation (3.13). These  involve the constants $N_A{}^B {}_C{}^D$ but these are determined by purely group theoretic considerations and are easy to compute at low levels, as indeed we have done in this paper for certain fields in section four. In section five we  gave a more explicit form for the gauge variations. 
\par
An interesting recent paper [30] derived the generalised  tangent space required for the closure of the generalised diffeomorphisms associated with U duality. They found  a hierarchy of tensor coordinates  which were those found in papers on $E_{11}$.  However, unlike some other recent works, reference [30] did not borrow the tangent space structures from $E_{11}$ papers but derived them from a different perspective.   The generalised tangent space encoded in the $E_{11}\otimes_s l_1$ non-linear realisation arises in a simple way from the $l_1$ realisation and it is the agreement with this that has been  found in reference [30]. The results contained in  [30] are also consistent with the approach of this paper that  takes the gauge parameters 
to belong to the $l_1$ representation. It would be interesting to make  detailed contact between this paper, reference [30] and other recent papers that consider local transformations associated with U duality. 
\par
As we have mentioned the equations of motion for the $E_{11}\otimes_s l_1$ non-linear realisation, when computed,  at low levels,  appear to be unique once one takes into account the symmetries encoded in the non-linear realisation at sufficiently high level. This is in contrast with the more common situation for more traditional non-linear realisations. As such it would be very interesting to see if the equations of motion that follow from the 
$E_{11}\otimes_s l_1$ non-linear realisation are invariant under the gauge transformations proposed in this paper. We have considered some of the very lowest level contributions to the equations of motion in four and five dimensions and found that they are indeed invariant under the proposed gauge transformations at the level considered. This is an encouraging sign. As we noted in section five,  a specific gauge transformation leads to the rigid $E_{11}$ transformations at least up to linear order in the fields. This will go some way to ensuring that the field equations are indeed gauge invariant. A similar remark applies to the closure of two gauge transformations. 
\par
The equations of motion for the $E_{11}\otimes_s l_1$ realisation have been computed to higher levels in fields and coordinates than those given in references [20] and [21], including a derivation of the equation of motion in   the gravity sector [31]. These results will also be useful for the exploration of the gauge transformations proposed in this paper. 
\par
We note that the construction of the equations of motion of the 
$E_{11}\otimes_s l_1$ non-linear realisation does not appear to demand that the fields are subject to some kind of condition, such as  the section condition that first appeared in Siegel theory [17]. The section condition does not appear to have a clear physical motivation and it is unclear if it allows for some of the theories that are known to exist. Given  that the familiar gauge parameters arise from generalised coordinates that are 
beyond those at levels zero and one,  it would seem natural to take the fields to also depend on the  coordinates of the generalised space-time which are at a higher level. This is indeed what happened when all the five dimensional maximal supergravities were constructed from the $E_{11}\otimes_s l_1$ non-linear realisation in reference [9]. However, the dependence of the fields  on the generalised coordinates was in a prescribed manner  from the outset.   It should be straightforward to recover this result by computing  the general  five dimensional equations of motion in which the way the fields depend on  the generalised coordinates 
is not prescribed and then inserting into these the required field dependence rather than working with this dependence from the beginning. 
In doing this calculation one will be able to see how the gauge transformations proposed in this paper match up with the local symmetries of the gauged supergravity and it would give an insight into how to restrict the dependence of the fields on the generalised space-time. This applies to any dimension and in particular to four dimensions where the equations of the $E_{11}\otimes_s l_1$ non-linear realisation are known [21,31] at low levels. 

\medskip
{\bf {Acknowledgment}}
\medskip 
I wish to thank Axel Kleinschmidt, Alexander Tumanov and Nikolay Gromov for discussions   and the SFTC for support from Consolidated grant number ST/J002798/1. 

{\centerline {\bf Appendix A}}
\medskip

In this appendix we give the commutators of the $E_{11}\otimes_s l_1$ algebra decomposed to $GL(5)\otimes E_6$, that is, the algebra relevant to the five dimensional theory. We used this algebra in section 4.1. The commutators involving the positive level $E_{11}$ generators were given in   references [9] and [29] and we use the conventions of the latter.  The techniques we use to derive this algebra can be used to find the $E_{11}\otimes_s l_1$ algebra in any other dimension. The commutation relations for the $E_6$ generators can be written as 
$$
  [R^\alpha , R^\beta ]= f^{\alpha\beta}{}_{\gamma} R^\gamma 
 \eqno(A.1)$$
where $f^{\alpha\beta}{}_{\gamma}$ are the structure constants of
$E_6$. The other generators  belong to specific representations of $E_6$ and this determines their commutators with the $E_6$ generators. Looking at the generators listed in equation (4.1.1) we find that 
$$
   [R^\alpha , R^{aM} ]= (D^\alpha )_N{}^M R^{a N} ,\quad 
   [R^\alpha , R^{ab}{}_M ]= -(D^\alpha )_M{}^N R^{ab}{}_N ,\ 
  \quad 
$$
$$
   [R^\alpha , R^{abc ,\beta} ]= f^{\alpha\beta}{}_{\gamma}
  R^{abc,\gamma} ,\quad 
   [R^\alpha , R^{abcd}{}_{MN} ]= -(D^\alpha )_M{}^P 
  R^{abcd}{}_{PN}-(D^\alpha )_N{}^P  R^{abcd}{}_{MP}\quad 
 \eqno(A.2)$$
where $(D^\alpha )_N{}^M $ obey the relation 
$$
  [D^\alpha , D^\beta ]_M{}^N = f^{\alpha\beta}{}_\gamma (D^\gamma )_M{}^N 
\eqno(A.3)$$
Here $f^{\alpha \beta \gamma}$ are the structure constants of
$E_6$ and are normalised so as to obey the relation. 
$$
  f_{\alpha \beta\gamma} f^{\alpha \beta \delta} = - 4
  \delta^\delta_\gamma
\eqno(A.4)$$
We lower and raise indices on the $E_6$ generators with the Cartan-Killing metric $g_{\alpha\beta}$ of $E_6$. The above matrices are normalised so that $$
  (D^\alpha)_M{}^N (D^\beta)_N{}^M = g^{\alpha \beta}
\eqno(A.5)$$ 
\par
The commutation relations of  the positive level generators are given by 
$$
 [R^{a M} , R^{b N} ]= d^{MNP} R^{ab}{}_P, \quad
 [R^{a N} , R^{bc}{}_M ]= g_{\alpha\beta} (D^\alpha )_M{}^N R^{abc, \beta} 
$$
$$
[R^{ab}{}_M , R^{cd}{}_N ]= R^{abcd}{}_{MN}, \quad
   [R^{a  P}, R^{bcd , \alpha} ]= S^{\alpha P, MN} R^{abcd}{}_{MN}
\eqno(A.6)$$
where $d^{MNP}$ is the completely symmetric invariant tensor of
$E_6$ formed from the product of three $\overline {\bf 27}$ representations. The invariant tensor $d_{MNP}$, which has its indices down,  is 
completely symmetric product of three ${\bf 27}$ indices and 
satisfies the relation 
$$
  d^{MNP} d_{MNQ} = \delta^P_Q 
\eqno(A.7)$$
We follow the conventions of reference [29] rather than reference [9], the difference being a rescaling of $d$ by $\sqrt{5}$. 
\par
The symbol $S^{\alpha P, MN}$, in the last of the  equations (A.6), 
 is  an invariant tensor, antisymmetric
with respect to $MN$, and the Jacobi identity between two 1-forms
and one 2-form implies the relation 
$$
  g_{\alpha \beta} D^\alpha_Q{}^{(P} S^{\beta R ) , MN} =  -{1 \over 2} \delta_{Q}^{[M} d^{N]PR}
\eqno(A.8)$$
Using the fact that $d^{MNP}$ is completely symmetric in its three indices one can derive from this the condition
$$
  g_{\alpha \beta} D^\alpha_M{}^N S^{\beta M , PQ} = 0 \quad .
\eqno(A.9)$$
We also have the relation [29] 
$$
  g_{\alpha \beta} D^\alpha_M{}^N D^\beta_P{}^Q = {1 \over 6} \delta^N_P
  \delta_M^Q + {1 \over 18} \delta_M^N \delta_P^Q -{5 \over 3} d^{NQR} d_{MPR}
\eqno(A.10)$$
from which one can  show the relations [29] 
$$
  S^{\alpha M , NP} = -3 D^\alpha_Q{}^{[N} d^{P] MQ} \quad
\eqno(A.11)$$
and  
$$
  S^{\alpha M ,NP} + {3 \over 2} ( D^\alpha D_\beta )_Q{}^M S^{\beta
  Q , NP} = 0\quad . 
\eqno(A.12)$$
\par
The commutation relations of equation (A.6) follow by writing down the most general possible commutation relations for the known $E_{11}$ generators so as to preserve the level and then enforcing the  Jacobi identities,  starting with the Jacobi identities involving one of  the $E_6$ generators.  
\par
The commutation relations between the generators of $E_6$ and the negative level generators are 
$$
 [R^\alpha , R_{a,M} ]= -(D^\alpha )_M{}^N R_{a, N} ,\quad 
   [R^\alpha , R_{ab}{}^N ]= (D^\alpha )_M{}^N R_{ab}{}^M ,
$$
$$
[R^\alpha , R_{abc , }{}^{ \beta} ]= f^{\alpha\beta}{}_{\gamma}
  R_{abc , }{}^{\gamma} ,\quad 
[ R^\alpha , R_{abcd}{}^{MN} ]= (D^\alpha )_P{}^M  
  R^{abcd}{}^{PN}+(D^\alpha )_P{}^N  R_{abcd}{}^{MP}\quad ,
\eqno(A.13)$$
\par
The Cartan involution acts on the generators of $E_{11}$ as follows 
$$
I_c(K^a{}_b)=- K^b{}_a ,\quad I_c(R^\alpha)=- R^{-\alpha}  ,\quad I_c(R^{aN})=- R_{aM} J^{MN} ,\quad
$$
$$
 I_c(R^{ab }{}_M)= J^{-1}_{MN} R_{ab}{}^N ,\quad I_c(R^{abc \alpha})= - R_{abc -\alpha},\quad I_c(R^{abcd}{}_{ MN})= J^{-1}_{MP} J^{-1}_{NQ} R_{abcd }{}^{PQ},
\eqno(A.14)$$
This equation follows from the knowledge that the Cartan involution takes positive level roots to negative level roots and  as a result it must  take superscript   space-time indices to  subscript   space-time indices. It also takes   a positive root $\alpha$   in $E_6$ to $-\alpha$ and,   as we have labelled the $E_6$ generators by their roots,  it acts on these as above. One could write a different matrix $J$ when the Cartan involution $I_c$ acts on each generator,  but it is easy to see that this will not lead to consistent commutators unless they are as given above. 
\par
Applying the Cartan involution  to the  commutators for the positive level $E_{11}$ generators,  given in equation (A.6), one finds the  commutators between the negative level $E_{11}$ generators:  
$$
 [R_{a M} , R_{b N} ]= d_{MNP} R_{ab}{}^P, \quad
 [R_{a N} , R_{bc}{}_M ]=  (D^\alpha )_N{}^M R_{abc, }{}_{\alpha} 
$$
$$
[R_{ab}{}^M , R_{cd}{}^N ]= R_{abcd}{}^{MN}, \quad
   [R_{a  P}, R_{bcd , \alpha} ]= S_{\alpha P, MN} R_{abcd}{}^{MN}
\eqno(A.15)$$
In carrying out this calculation one finds   the relations 
$$
f^{-\alpha -\beta}{}_{-\gamma}= - f^{\alpha \beta}{}_{\gamma},\quad 
J^{MP}(D^{-\alpha})_P{}^Q J^{-1}_{QN}=  (D^\alpha)_N{}^M ,\quad 
$$
$$
d_{PQR}= d^{MNS} J^{-1}_{MP}J^{-1}_{NQ}J^{-1}_{SR}, \quad 
S_{\alpha S, RQ}= S^{-\alpha P, MN} J^{-1}_{PS}J^{-1}_{MR}J^{-1}_{NQ}
\eqno(A.16)$$
\par
The commutators between the positive and negative level generators of $E_{11}$ are given by 
$$
[ R^{aN} , R_{bM} ]= 6\delta^a_b (D^\alpha)_M{}^N R^\alpha + \delta_M^N K^a{}_b -{1\over 3} \delta_M^N \delta^a_b \sum_c K^c{}_c
$$
$$
[ R_{aN} , R^{bc}{}_{M} ]= 20 d_{NMP} \delta_a^{[b} R^{c] P} , \quad 
[R_{aN} , R^{b_1b_2b_3 \alpha} ] =18 (D^\alpha)_N{}^M \delta _a^{[b_1}R^{b_2b_3 ]}{}_{M}
\eqno(A.17)$$

\par
We now give the commutators between the generators of $E_{11}$ and those of the $l_1$ representation. The commutation relations between the later and   the generators of GL(5) are given by 
$$
[K^a{}_b, P_c]= - \delta^a_cP_b+{1\over 2} \delta^a_bP_c,\quad  
[K^a{}_b, Z^N]= {1\over 2} \delta^a_b Z^N , \quad 
$$
$$
[K^a{}_b, Z_N^c]=  \delta^c_b Z^a_N +{1\over 2} \delta^a_bZ^c_N,\quad  
[K^a{}_b, Z^{c_1c_2\alpha}]=\delta^{c_1}_bZ^{ac_2 \alpha}
+\delta^{c_2}_bZ^{c_1 a \alpha} +{1\over 2} \delta^a_b Z^{c_1c_2\alpha}
\eqno(A.18)$$
while with the generators of $E_6$ we have 
$$
  [R^\alpha , P_a]=0,\ 
[R^\alpha , Z^M]= Z^N (D^\alpha)_N{}^M,\ [R^\alpha , Z^a_N]= - (D^\alpha)_N{}^M Z^a_M
$$
$$
  [R^\alpha , Z^{a_1a_2\beta}]=f^{\alpha\beta}{}_\gamma Z^{a_1a_2\gamma},\ [R^\alpha , Z^{a_1a_2a_3}{}_{NM}]=
  -(D^\alpha)_N{}^R Z^{a_1a_2a_3}_{RM} -(D^\alpha)_M{}^R Z^{a_1a_2a_3}{}_{NR} 
\eqno(A.19)$$
\par
The commutators of certain of the positive root generators of $E_{11}$ with the space-time translations can  be taken to be of the form
$$
[R^{aN}, P_b]=\delta_b^a Z^N,\ [R^{a_1a_2}_{N}, P_b]=2 \delta_{b}^{[a_1}Z^{a_2]}_N,
$$
$$
[R^{a_1a_2a_3\alpha}, P_b]=3 \delta_{b}^{[a_1}Z^{a_2a_3]\alpha},\  [R^{a_1a_2a_3a_4}{}_{MN}, P_b]=4
\delta_{b}^{[a_1}Z^{a_2a_3a_4]}{}_{MN} \ . 
\eqno(A.20)$$ 
The coefficients on the right-hand side can be  freely   chosen  and this fixes the normalisation of the generators that appear on this side of the commutators. The commutation relations of the remaining positive generators of $E_{11}$ follow form equation (A.20), the commutators of the $E_{11}$ algebra,  and the Jacobi relations. 
One finds that 
$$
  [R^{aM}, Z^N]= -d^{MNP} Z^a_P \quad 
[R^{aM}, Z^b_N]= -(D_\alpha)_N{}^M Z^{ab\alpha},\ 
$$
$$
[R^{a_1a_2}_{M}, Z^N]= -(D_\alpha)_M{}^N Z^{a_1a_2\alpha},\quad
[R^{a_1a_2}_{M}, Z^{a_3}_{N}]=Z^{a_1a_2a_3}_{MN},\ 
$$
$$
 [ R^{a_1M}, Z^{a_2a_3\alpha}]=-S^{\alpha M [RS]}
Z^{a_1a_2a_3}{}_{RS}, \quad
[R^{a_1a_2a_3\alpha}, Z^{M}]=-S^{\alpha M [ RS]} Z^{a_1a_2a_3}{}_{RS} \quad . \eqno(A.21)$$
\par
The commutators between the level -1 generators of $E_{11}$ and those of the $l_1$ representation are 
$$
[R_{aN} , Z^M]= \delta_N^M P_a , \quad 
[R_{aN} , Z_M^b ]= -10 d_{NMP} \delta_a^b Z^P , \quad
$$
$$
[R_{aN} , Z^{b_1b_2 \alpha}]= -{12} (D^\alpha)_N{}^P \delta_a^{[b_1}Z^{b_2]}_P
\eqno(A.22)$$


\medskip
{\bf {References}}
\medskip
\item{[1]} P. West, {\it $E_{11}$ and M Theory}, Class. Quant.  
Grav.  {\bf 18}
(2001) 4443, {\tt arXiv:hep-th/ 0104081}; 
\item{[2]} P. West, {\it $E_{11}$, SL(32) and Central Charges},
Phys. Lett. {\bf B 575} (2003) 333-342, {\tt hep-th/0307098}
\item{[3]}  A. Borisov and V. Ogievetsky,  {\it Theory of dynamical affine and conformal  symmetries as the theory of the gravitational field}, 
Teor. Mat. Fiz. 21 (1974) 32
\item{[4]} P. West, {\it Hidden superconformal symmetries of  
M-theory}, {\bf JHEP 0008} (2000) 007, {\tt arXiv:hep-th/0005270}.
\item{[5]} P. West, Introduction to Strings and Branes, Cambridge
University Press, June 2012. 
\item{[6]} I. Schnakenburg and  P. West, {\it Kac-Moody   
symmetries of
IIB supergravity}, Phys. Lett. {\bf B517} (2001) 421, {\tt  
arXiv:hep-th/0107181}.
\item{[7]} P. West, {\it The IIA, IIB and eleven dimensional theories 
and their common
$E_{11}$ origin}, Nucl. Phys. B693 (2004) 76-102, hep-th/0402140. 
\item{[8]}  F. ÊRiccioni and P. West, {\it
The $E_{11}$ origin of all maximal supergravities}, ÊJHEP {\bf 0707}
(2007) 063; ÊarXiv:0705.0752.
\item{[9]} ÊF. Riccioni and P. West, {\it E(11)-extended spacetime
and gauged supergravities},
JHEP {\bf 0802} (2008) 039, ÊarXiv:0712.1795.
\item{[10]} P. West,  {\it $E_{11}$ origin of Brane charges and U-duality
multiplets}, JHEP 0408 (2004) 052, hep-th/0406150. 
\item{[11]} P. West, {\it Brane dynamics, central charges and
$E_{11}$}, hep-th/0412336. 
\item{[12]} P. Cook and P. West, {\it Charge multiplets and masses
for E(11)}, ÊJHEP {\bf 11} (2008) 091, arXiv:0805.4451.
\item{[13]}  A. Kleinschmidt and P. West, {\it  Representations of G+++
and the role of space-time},  JHEP 0402 (2004) 033,  hep-th/0312247.
\item{[14]} E. Bergshoeff, I. De Baetselier and  T. Nutma, {\it 
E(11) and the Embedding Tensor},  JHEP 0709 (2007) 047, arXiv:0705.1304 
\item{[15]} M. Duff, {\it Duality Rotations In String Theory},
  Nucl.\ Phys.\  B {\bf 335} (1990) 610; M. Duff and J. Lu,
 Duality rotations in
membrane theory,  Nucl. Phys. {\bf B347} (1990) 394. 
\item{[16]} A. Tseytlin, {\it Duality Symmetric Formulation Of String World
Sheet Dynamics}, Phys.Lett. {\bf B242} (1990) 163, {\it Duality Symmetric
Closed String Theory And Interacting Chiral Scalars}, Nucl.\ Phys.\ B {\bf
350}, 395 (1991). 
\item{[17]} W. Siegel, {\it Two vielbein formalism for string inspired axionic gravity},   Phys.Rev. D47 (1993) 5453,  hep-th/9302036; 
	{\it Superspace duality in low-energy superstrings}, Phys.Rev. D48 (1993) 2826-2837, hep-th/9305073; 
{\it Manifest duality in low-energy superstrings},  
In *Berkeley 1993, Proceedings, Strings '93* 353,  hep-th/9308133. 
\item {[18]} C. Hillmann, {\it Generalized E(7(7)) coset dynamics and D=11
supergravity}, JHEP {\bf 0903}, 135 (2009), hep-th/0901.1581. 
\item{[19]} C. Hillmann, {\it E(7(7)) and d=11 supergravity }, PhD
thesis,  arXiv:0902.1509.
\item {[20]} P. West, {\it Generalised Geometry, eleven dimensions
and $E_{11}$}, JHEP 1202 (2012) 018, arXiv:1111.1642.  
\item{[21]} P. West, {\it  E11, Generalised space-time and equations of motion in four dimensions}, JHEP 1212 (2012) 068, arXiv:1206.7045. 
\item{[22]} O. Hohm and S.  Kwak, {\it Frame-like Geometry of Double Field Theory},   J.Phys.A44:085404,2011, arXiv:1011.4101. 
\item{[23]}  P. West, {\it E11, generalised space-time and IIA string
theory}, 
 Phys.Lett.B696 (2011) 403-409,   arXiv:1009.2624.
\item{[24]}   A. Rocen and P. West,  {\it E11, generalised space-time and
IIA string theory;  the R-R sector},  arXiv:1012.2744.
\item{[25]} O. Hohm and  H. Samtleben, {\it Exceptional Field Theory I: $E_{6(6)}$ covariant Form of M-Theory and Type IIB}, 
\item{[26]}  O. Hohm and  H.  Samtleben, {\it Exceptional Field Theory II: E$_{7(7)}$} ,  arXiv:1312.0614. 
\item{[27]} A. Coimbra, C. Strickland-Constable and  D.  Waldram, 
{\it Supergravity as Generalised Geometry I: Type II Theories}, 
arXiv:1107.1733; {\it $E_{d(d)} \times {R}^+$ Generalised Geometry,
Connections and M theory}, arXiv:1112.3989. A. Coimbra, C. Strickland-Constable and  D.  Waldram, {\it  Supergravity as Generalised Geometry II: $E_{d(d)} \times {R}^+$ and M theory} , arXiv:1212.1586; P. Pacheco and  D.  Waldram, {\it M-theory, exceptional generalised geometry and superpotentials},  JHEP0809 (2008) 123,  arXiv:0804.1362 
\item{[28]} N. Hitchin, Generalized Calabi-Yau manifolds, 
 Q. J. Math.  {\bf 54}  (2003), no. 3, 281,
math.DG/0209099;  {\it Brackets, form and
invariant functionals}, math.DG/0508618;  M. Gualtieri, {\it Generalized complex geometry}, PhD Thesis
(2004), math.DG/0401221v1. 
\item{[29]} F. ÊRiccioni, ÊD. ÊSteele and P. West, {\it The E(11)
origin of all maximal supergravities - the hierarchy of field-strengths}
ÊÊJHEP {\bf 0909} (2009) 095, arXiv:0906.1177. 
\item{[30]} G.  Aldazabal, M. Grana, D.  MarquŽs, and J. Rosabalar, 
{\it The gauge structure of Exceptional Field Theories and the tensor hierarchy}, Xiv:1312.4549. 
\item{[31]} N. Gromov and P. West, to be published. 
     
\item{[32]} D. Berman, H. Godazgar, M. Perry and P. West, {\it Duality Invariant Actions and Generalised Geometry}, JHEP 1202 (2012) 108, arXiv:1111.0459 

\end